\documentclass[aps,pre,superscriptaddress,groupedaddress]{revtex4}  % for review and submission

\usepackage{graphicx}  % needed for figures
\usepackage{placeins}
\usepackage{url}                  %this package should fix any errors with URLs in refs.
\usepackage{dcolumn}       % needed for some tables
\usepackage{bm}                  % for math
\usepackage{amssymb}       % for math
\usepackage{mathtools}     % for math
\usepackage{booktabs}
\usepackage{amsmath}    % mathematical env.
\usepackage{pgf}                   %plots from matplotlib in TeX style
\usepackage{tikz}                   %draw graphics
\usepackage{lipsum}
\usepackage{algorithm}% PseudoCode
\usepackage{algpseudocode}
\usepackage{float}

\usepackage{caption}

\begin{document}

\title{Generative convective parametrization of dry atmospheric boundary layer}

\author{Florian Heyder}
\affiliation{Institut f\"ur Thermo- und Fluiddynamik, Technische Universit\"at Ilmenau, Postfach 100565, D-98684 Ilmenau, Germany}

\author{Juan Pedro Mellado}
\affiliation{Meteorologisches Institut, Universit\"at Hamburg, Bundesstra\ss e 55, D-20146 Hamburg, Germany}

\author{J\"org Schumacher}
\affiliation{Institut f\"ur Thermo- und Fluiddynamik, Technische Universit\"at Ilmenau, Postfach 100565, D-98684 Ilmenau, Germany}
\affiliation{Tandon School of Engineering, New York University, New York City, NY 11201, USA}

\date{\today}

\begin{abstract}
Turbulence parametrizations will remain a necessary building block in kilometer-scale Earth system models. In convective boundary layers, where the mean vertical gradients of conserved properties such as potential temperature and moisture are approximately zero, the standard ansatz which relates turbulent fluxes to mean vertical gradients via an eddy diffusivity has to be extended by mass flux parametrizations for the typically asymmetric up- and downdrafts in the atmospheric boundary layer. In this work, we present a parametrization for a dry convective boundary layer based on a generative adversarial network. The model incorporates the physics of self-similar layer growth following from the classical mixed layer theory by Deardorff. This enhances the training data base of the generative machine learning algorithm and thus  significantly improves the predicted statistics of the synthetically generated turbulence fields at different heights inside the boundary layer. The algorithm training is based on fully three-dimensional direct numerical simulation data. Differently to stochastic parametrizations, our model is able to predict the highly non-Gaussian transient statistics of buoyancy fluctuations, vertical velocity, and buoyancy flux at different heights thus also capturing the fastest thermals penetrating into the stabilized top region. The results of our generative algorithm agree with standard two-equation or multi-plume stochastic mass-flux schemes. The present parametrization provides additionally the granule-type horizontal organization of the turbulent convection which cannot be obtained in any of the other model closures. Our work paves the way to efficient data-driven convective parametrizations in other natural flows, such as moist convection, upper ocean mixing, or convection in stellar interiors.   
\end{abstract}

\maketitle

\section{Introduction}
Convective phenomena occur at various spatial and temporal scales in the atmosphere, ranging from meter-scale thermals at the surface \cite{Lenschow1980} via individual clouds or cloud layers \cite{Bodenschatz2010,Mellado2017} and mesoscale convection \cite{Atkinson1996,Bony2020} to global-scale patterns, such as the monthly timescale Madden-Julian oscillation in the tropics \cite{Madden1972,Miura2007}. Capturing this spatial heterogeneity, the transition between different convective regimes, and the resulting fluxes remains a challenge and requires turbulence parametrizations of the dynamics at the scales which remain unresolved in the numerical models. Mass-flux parametrizations are one specific way of representing unresolved convective sub-grid scale processes in atmospheric turbulence \cite{Ertel1942,Deardorff1972,Arakawa1974}, see also \cite{LeMone:2019}; they belong to the class of counter-gradient models which become necessary in the absence of a mean gradient of conserved properties \cite{Tiedtke1989,Siebesma1995,Zeli2021}. These parametrizations of the dynamics in atmospheric boundary layers (ABL) capture the effective vertical transport of heat, moisture, and momentum in the lower atmosphere in a set of reduced equations that are solved together with the transport equations for energy and total water content \cite{Wyngaard:2010}. By incorporating mass-flux schemes, kilometer-scale global circulation models can couple the complex interactions between large-scale dynamics and smaller-scale convective processes at mesoscales, essential for improved predictions of shallow cloud patterns \cite{Klingebiel2021,Vogel2022} and resulting rainfall distributions \cite{Ahlgrimm2014,Becker2021} or short-term climate change prognoses \cite{Hohenegger2023}. While these physical parametrization models have proven to be valuable tools, they have some limitations resulting from a set of various free parameters that have to be adjusted to control the physical processes \cite{Siebesma2007}. Furthermore, they do not provide any information on the horizontal organisation of the updrafts which determines the cloud bases that form on the top of the dry layer. 

In the past years, machine learning (ML) methods have shown promise as an alternative and data-driven approach to model convective processes even though their generalization capabilities remain often limited \cite{Brenowitz2018,Reichstein2019,Kashinath2021}. These algorithms have been applied to classify convection patterns \cite{Fonda2019, Rasp2020}; they served as subgrid-scale models in large-eddy simulations \cite{BeckKurzGAMM2021, Yuval2021, Cheng2022, Guan2023}, learned the nudging tendencies for the correction of physical parametrizations \cite{Bretherton2022}, or were used as reduced-order models in the form of recurrent neural networks to reproduce low-order turbulence statistics \cite{Heyder2021,Heyder2022}. Deep generative ML algorithms, e.g., in the form of Generative Adversarial Networks (GANs), create synthetic data which correspond to convection phenomena, such as cloud cover \cite{Dai2021}, vertical cloud structure \cite{Leinonen2019}, or precipitation distributions \cite{Price2022}. Shamekh and Gentine used a variational autoencoder, a further generative machine learning model, to approximate the mean vertical profiles of the buoyancy flux due to wind shear and convection \cite{Shamekh2023}. This specific class of machine learning methods opens thus doors to access the horizontal organisation of the turbulent convection and sets the stage and motivation for the present investigation.

In this work, we present a data-driven mass-flux parametrization of the buoyancy flux in the form of a GAN \cite{Goodfellow2014,Goodfellow-et-al-2016}. We therefore consider a paradigm for the dry atmospheric boundary layer, a shear-free flux-driven convective boundary layer. In a high-fidelity three-dimensional direct numerical simulation (DNS), we fully resolve the transient growth of the ABL \cite{Mellado2012,Garcia2014}. The DNS snaphots provide training data for our semi-supervised ML method \cite{Brunton2020}.  The GAN architecture consists of two competing deep neural networks; (1) the generator is here a U-shaped convolutional neural network, in short U-Net \cite{Ronneberger2015}, (2) the discriminator is a deep neural network. The generator models the {\em transient growth} of the ABL by inclusion of the physical laws of self-similar growth of the layer \cite{Deardorff1972}. In this way, we augment the training data base and thus enhance the prediction capability of the ML method significantly. The GAN is set up for different heights in the ABL, it provides the complete non-Gaussian statistics of the vertical velocity, buoyancy fluctuations, and buoyancy flux at reference heights in the sub-grid scale range. Our model is also capable to generate synthetic data with the right area fraction of updraft regions. It is found to agree with a mass-flux parametrization in the form of a set of reduced equations with adjustable parameters. The latter is for example applied for the atmospheric boundary layer in the Integrated Forecasting System, a global circulation model of European Centre for Medium-Range Weather Forecasts \cite{Siebesma2007,Witte2022}.   

%--------------------------------------------------
\begin{figure}[!htpb]
\centering
\includegraphics[width=\textwidth]{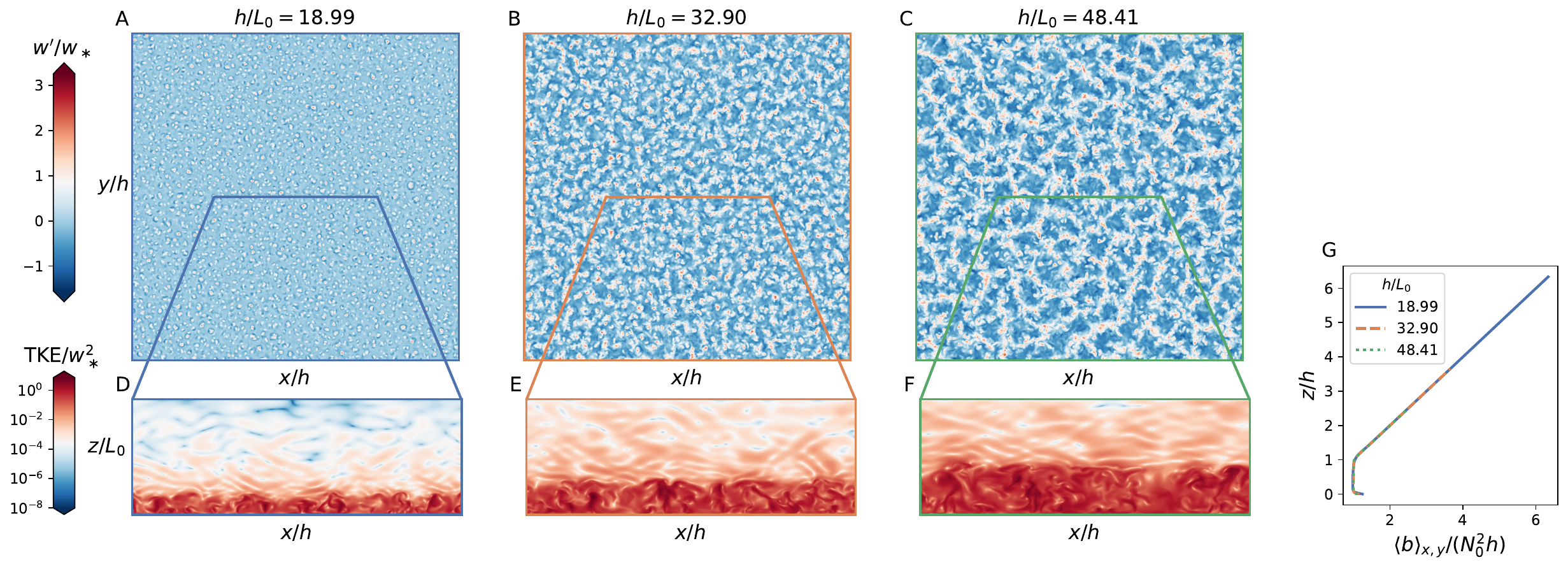}
\caption{Self-similar temporal growth of the atmospheric boundary layer. (A-C) Contours of the vertical velocity fluctuations $w^{\prime}$ viewed from the top onto the whole horizontal domain for three different times. (D-F) Corresponding turbulent kinetic energy, ${\rm TKE}=(u^{\prime\,2}+v^{\prime\,2}+w^{\prime\,2})/2$, in a vertical contour plot. The horizontal extension of the cut is indicated in (A-C). (G) Plane averaged vertical profiles of the buoyancy versus height at three time instants from panels (A-C).}
\label{fig:Fig1}
\end{figure}
%--------------------------------------------------
\section{Convective boundary layer}
We solve the three-dimensional Boussinesq equations which couple the two turbulent fields, the velocity field ${\bf v}=(u,v,w)$ and the buoyancy field $b$, in DNS that resolve all physically relevant scales down to the dissipation scales. In an atmospheric boundary layer without phase changes, the buoyancy is directly proportional to the virtual potential temperature field $\theta_\mathrm{v}$ and given by
\begin{equation}
b({\bf x},t)=g \frac{\theta_\mathrm{v}({\bf x},t)-\theta_\mathrm{v,0}}{\theta_\mathrm{v,0}}\,,
\label{eq:buo}
\end{equation}
with the acceleration due to gravity $g$ and a reference temperature $\theta_\mathrm{v,0}$. Convection proceeds in a transiently and self-similarly growing convective boundary layer (CBL) with an encroachment height $h(t)$ that is to a good approximation given by \cite{Deardorff1972,Garcia2014}
\begin{equation}
h(t)=L_0 \left[2(1+Re_0^{-1})N_0(t-t_0)\right]^{1/2}\,.
\label{eq:growth}
\end{equation}
This height contains the Brunt-V\"ais\"al\"a frequency $N_0$, the reference time $t_0$, the Reynolds number $Re_0$, and the reference Ozimidov-type scale $L_0=(B_0/N_0^3)^{1/2}$. It is driven by prescribed buoyancy fluxes at the bottom and top of the layer. The prescribed constant flux at the bottom is $B_\mathrm{v,0}$. Consequently, the convective Rayleigh number $Ra_c(t)=B_0 h(t)^4/(\nu\kappa^2)$ grows with time and reaches at an intermediate time a value of $3\times 10^8$ in the present case. Here, $\nu$ and $\kappa$ are the kinematic viscosity and buoyancy diffusivity, respectively. The Prandtl number is $Pr=\nu/\kappa=1$. Further details on the model equations, the essential dimensionless parameters, and the numerical simulation technique are provided in the {\color{blue} SI Appendix}.      

The buoyancy flux is the central transport measure. It couples buoyancy and vertical velocity fluctuations, i.e., the deviations from the respective vertical mean profiles of both fields. In an eddy-diffusivity mass-flux approach, it is modeled as follows,
\begin{equation}
\rho_0\langle w^{\prime}b^{\prime}(z)\rangle_{x,y} = -\rho_0\kappa_t \frac{\partial \langle b\rangle_{x,y}}{\partial z} + M_u (\langle b(z)\rangle_u-\langle b(z)\rangle_{x,y})\,,
\label{eq:mass1}
\end{equation}
where $\rho_0$ is the constant reference density in the convective boundary layer and $M_u$ the mass flux parameter. Here, $\langle\cdot\rangle_{x,y}$ are horizontal plane averages. The first term in \eqref{eq:mass1} is the standard Boussinesq term for a turbulent stress which contains the turbulent diffusivity $\kappa_t$. The second term on the right hand side stands for the mass-flux parametrization which has to be included when the mean buoyancy gradient vanishes for the well-mixed layer, as being the case here. 

\section{Up -and downdraft regions}

Figure \ref{fig:Fig1} illustrates the temporal growth of the convective boundary layer. While panels (A--C) provide a view from the top for the vertical velocity fluctuations with an aggregation of the convection cells, panels (D--F) display the turbulent kinetic energy in a vertical cut illustrating the transient growth  process. The remaining panel (G) of Fig. \ref{fig:Fig1} demonstrates that this growth follows a self-similar process and a rescaling of lengths with $h(t)$ and buoyancy with $N_0^2h(t)$ collapses the vertical mean buoyancy profiles at different times, $\langle b(z,t)\rangle_{x,y}$. It is exactly this self-similar behaviour given by \eqref{eq:growth} for the growth of the mesoscale patterns in vertical and horizontal directions which will be used to augment the training data base for the generative network. 
%--------------------------------------------------
\begin{figure}[!htpb]
\centering
\includegraphics[width=0.4\textwidth]{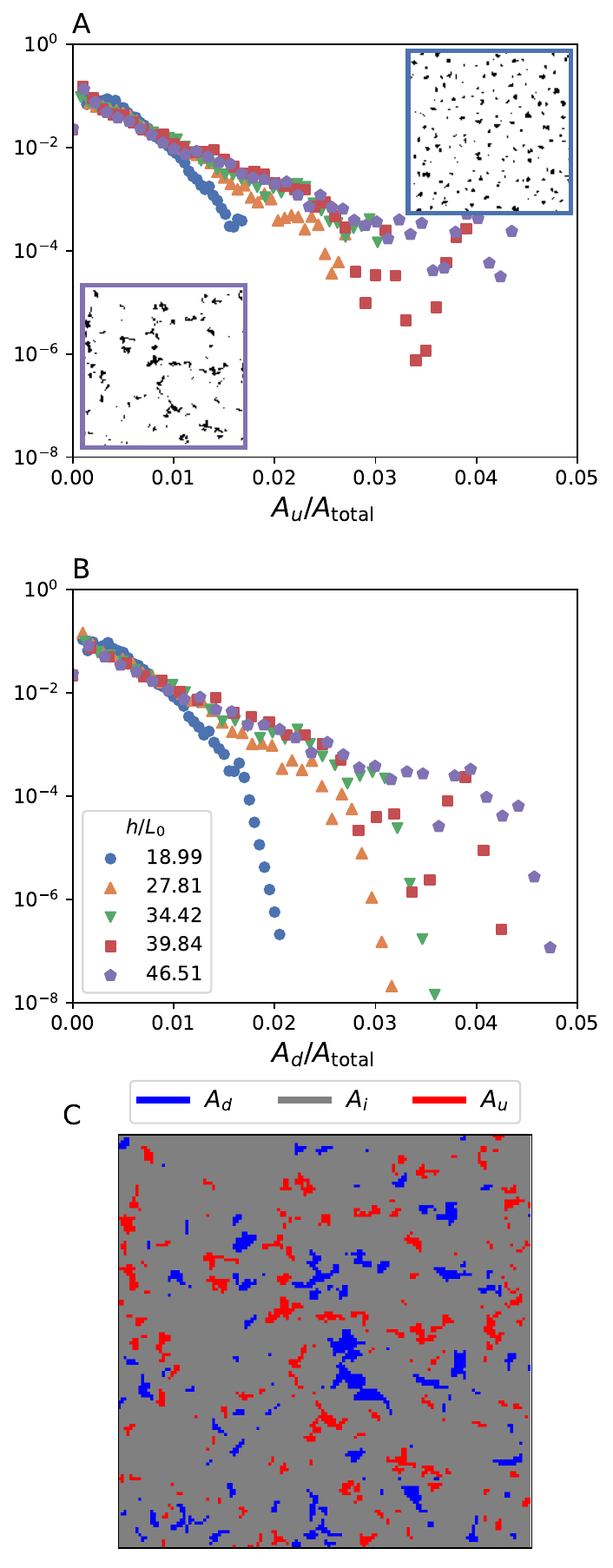}
\caption{Statistics of the up- and down-welling fluid motion in the convective boundary layer (top $95\%$ of $w$-values). (A) Probability density functions (PDFs) of the area $A_u$ of the up-welling fluid motion normalized by the total horizontal area $A_{\rm total}$ as a function of $h/L_0$ (see color legend in (B)). The two insets display a cutout of the updrafts in a fixed horizontal plane $z=0.5$ at different times. (B) PDF of the area $A_d$ of the down-welling fluid motion normalized by the total horizontal area $A_{\rm total}$ as a function of time. (C) Horizontal cutout which explains the decomposition of $A_{\rm total}$ into regions of strongest updrafts $A_u$, strongest downdrafts $A_d$, and the intermediate motions $A_i$. Note that the union of the regions of the strongest downdrafts and intermediate motions yields the environment $A_e$ for the mass-flux formulation.}
\label{fig:Fig2}
\end{figure}
%--------------------------------------------------

The prefactor $M_u$ in the second term of \eqref{eq:mass1} incorporates the specifics of the upward transport. As detailed in {\color{blue}SI Appendix}, the unknown mass-flux parameter $M_u$ is estimated as a product of the updraft area fraction, $a_u$, and the excess of the mean vertical velocity in the updraft regions, $\langle w\rangle_u$, over the mean upward motion, i.e., 
\begin{equation}
M_u\approx \rho_0 a_u (\langle w\rangle_u-\langle w\rangle_{x,y})\,.
\label{eq:mass2}
\end{equation} 
The horizontal cross section of the layer $A_{\rm total}$ is decomposed into disjoint strongest updraft (u) and downdraft (d) regions, as well as the remaining intermediate motions region (i), i.e., $A_{\rm total}=A_u \cup A_d \cup A_i$. We have thus further refined the environment region, which is typically taken as $A_e = A_i \cup A_d$, e.g., when the mass-flux models are discussed. This step is taken to analyse the asymmetry between up-and down-welling motion of same strength.  While $\langle\cdot\rangle_{x,y}$ are horizontal averages with respect to the whole cross section $A_{\rm total}$, $\langle\cdot\rangle_{u}$ denotes an average over the updraft and regions, see also \eqref{eq:mass2}. 

Figure \ref{fig:Fig2} underlines our observation from Fig. \ref{fig:Fig1} more quantitatively. The buoyancy flux is arranged in narrower updraft and somewhat broader downdraft regions in the convective boundary layer. Here, we show the probability density functions (PDFs) for both cases at different times of the transient growth. A direct comparison of the top and bottom panels shows that the tails of the downdraft distribution have somewhat broader tails. Consequently, downdrafts occupy slightly broader areas which underlines the top-down asymmetry of the convective motion. Moreover, it can be seen that the PDFs collapse increasingly better on each other in the core and parts of the tails as $h/L_0$ grows. The insets in the Fig. \ref{fig:Fig1}(A) underline that the cell width of the updraft patterns grows with time. This indicates the regime, where the boundary layer growth rate becomes small compared to the turbulent mixing rate by the convective motion.
%--------------------------------------------------
\begin{figure}[t!]
\centering
\includegraphics[width=0.5\columnwidth]{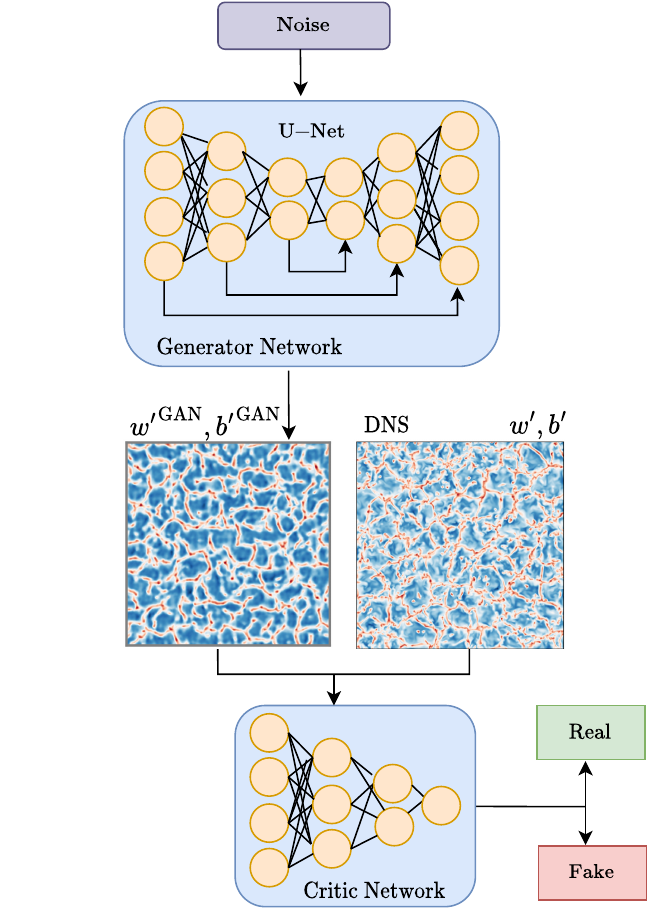}
\caption{Sketch of the architecture of the applied generative adversarial network. The generator is composed of a U-shaped deep network, a U-Net. Details, training procedure, and Wasserstein loss metrics are detailed in the SI text.}
\label{fig:Fig3}
\end{figure}
%---------------------------------------------------
\section{Generative machine learning model}
\subsection{Architecture} We introduce in brief the generative ML model as well the augmentation of the training database. Figure \ref{fig:Fig3} illustrates the concept of the generative adversarial network approach: two convolutional neural networks compete in an adversarial process and reach a final equilibrium state, similar to the Nash equilibrium in a two-player non-cooperative game \cite{Nash1950}. Here, the generator network has the architecture of a U-Net~\cite{Ronneberger2015}, which applies additional residual connections between the contraction and expansion paths of the network and is used for complex segmentation tasks with few training examples~\cite{Ronneberger2015,Fonda2019}. It is seeded by random latent variables and outputs a horizontal slice of fluctuations of vertical velocity ${w'}^{\rm GAN}(x,y)$ and buoyancy ${b'}^{\rm GAN}(x,y)$. The second network, the discriminator or critic, receives both the synthetic snapshot and a true DNS snapshot ${w'}(x,y)$, ${b'}(x,y)$. It then emits a score indicating whether the current input snapshot stems from the DNS, labeled as real, or the generator, labeled as fake. Finally, both networks are trained via stochastic gradient descent. While the generator is trained to fool the critic network, the latter aims to classify both snapshots correctly. 
%---------------------------------------------------
\begin{figure}[!htpb]
\centering
\includegraphics[width=0.83\textwidth]{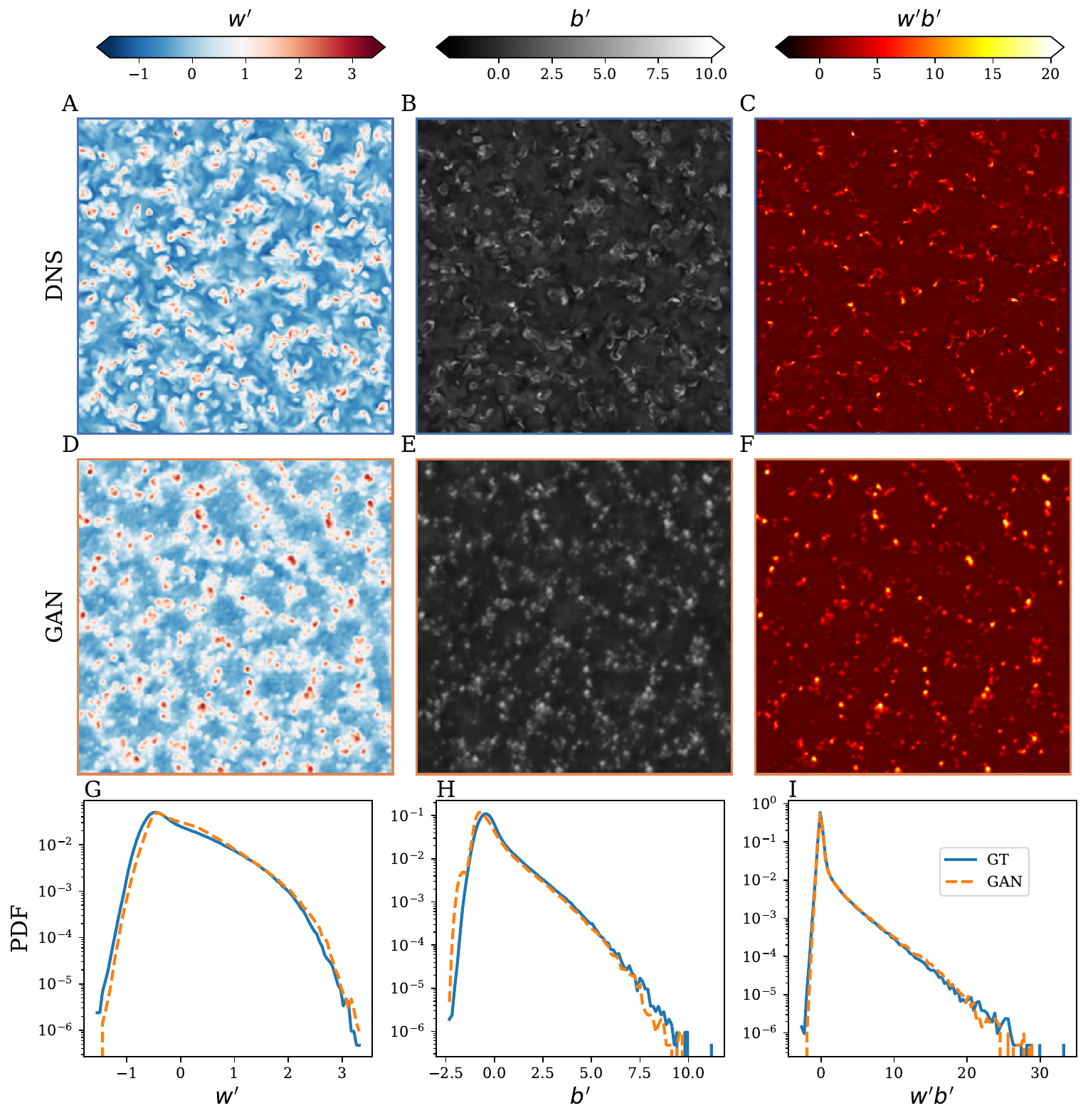}
\caption{Comparison of the generated turbulent fields with the ground truth. (A--C) Horizontal cross sections at $z=0.5h=0.584$ at $h/L_0=48.41$ of the vertical velocity fluctuations $w^{\prime}$ (A), the buoyancy fluctuations $b^{\prime}$ (B), and the local convective heat flux $w^{\prime}b^{\prime}$ (C) from DNS which is the ground truth (GT). (D--F) Corresponding cross sections which have been generated by the generative adversarial network (GAN). (G-I) Comparison of the probability density functions of the fields. }
\label{fig:Fig4}
\end{figure}
%-----------------------------------------------------

Here, we use the Wasserstein--GAN formulation introduced by Arjovsky \textit{et al.} \cite{Arjovsky2017} which applies a specific loss function for the optimization of the networks. The Wasserstein loss helps to overcome an initial training phase faster, where the generator network outputs sub-optimal snapshots, which are easily identified by the critic. Furthermore, it increases the overall stability of the network training process. Details on the Wasserstein--GAN implementation and the training can be found in the {\color{blue} SI Appendix}. After the successful training, one uses the optimized generator network only to produce synthetic snapshots of $w'$ and $b'$ at a given height and for different times without knowing the underlying Boussinesq equations. In this way, we directly obtain the local buoyancy flux field, $w'b'(x,y,z_0,t)$, at a given height $z_0$ inside the CBL and hence its horizontal average $\langle w'b'\rangle_{x,y}$ without the need of a physical parametrization model as described in (\ref{eq:mass1}). More importantly, this data-driven convective parametrization will be generalizable, i.e., incorporating the transient dynamics of the CBL since the network is trained with data that are rescaled in correspondence with the self-similar mixing layer theory which we detail in the following.

\subsection{Physics-informed data augmentation} We augment the training data in a re-normalization procedure that removes the major first-order effects of the CBL growth. For this, we use snapshots in the statistical quasi-steady phase of the CBL, where the spatial patterns of the up- and downdrafts are not affected by the very initial fast CBL growth, see again Fig. \ref{fig:Fig2}. In this regime, a new set of variables can be introduced that embeds, to leading order, the transient temporal growth of the boundary layer height. To this end, we rescale the turbulence fields with the similarity variables which correspond to the mixing layer theory by Deardorff \cite{Deardorff:1970}. This gives the following transformation rules,
\begin{equation}
\tilde{w}(\tilde{\bf x},\tilde{h}(t)) = \frac{w({\bf x},t)}{(B_0 h(t))^{1/3}}\,,\;\;\;\; 
\tilde{b}(\tilde{\bf x},\tilde{h}(t)) = \frac{b({\bf x},t)}{(B_0^2/h(t))^{1/3}}\,,
\end{equation}
with rescaled time $\tilde{h}(t)=h(t)/L_0$ and the normalized coordinates $\tilde{\bf x}={\bf x}/h(t)$. Note that this transformation leads to a decreasing horizontal extent as the boundary layer grows. The training data of the GAN is then chosen to be a horizontal slice at a plane $z/h(t)$ with constant horizontal extent in normalized coordinates for all snapshots. The latter step then involves a cropping procedure for most snapshots, which in turn results in similar-sized spatial patterns of $\tilde{w}$ and $\tilde{b}$ and thus of $\tilde{w}^{\prime}\tilde{b}^{\prime}$. In summary, this results in an augmented training database that encodes the effects of a growing boundary layer to the first order. The similar spatial features are then easily picked up by the generator network. Finally, we reverse the re-normalization procedure to obtain synthetic snapshots at any given rescaled time $h/L_0$. Further details on the augmentation are found in {\color{blue}SI Appendix}.
%---------------------------------------------------
\begin{figure}[!htpb]
\centering
\includegraphics[width=0.95\textwidth]{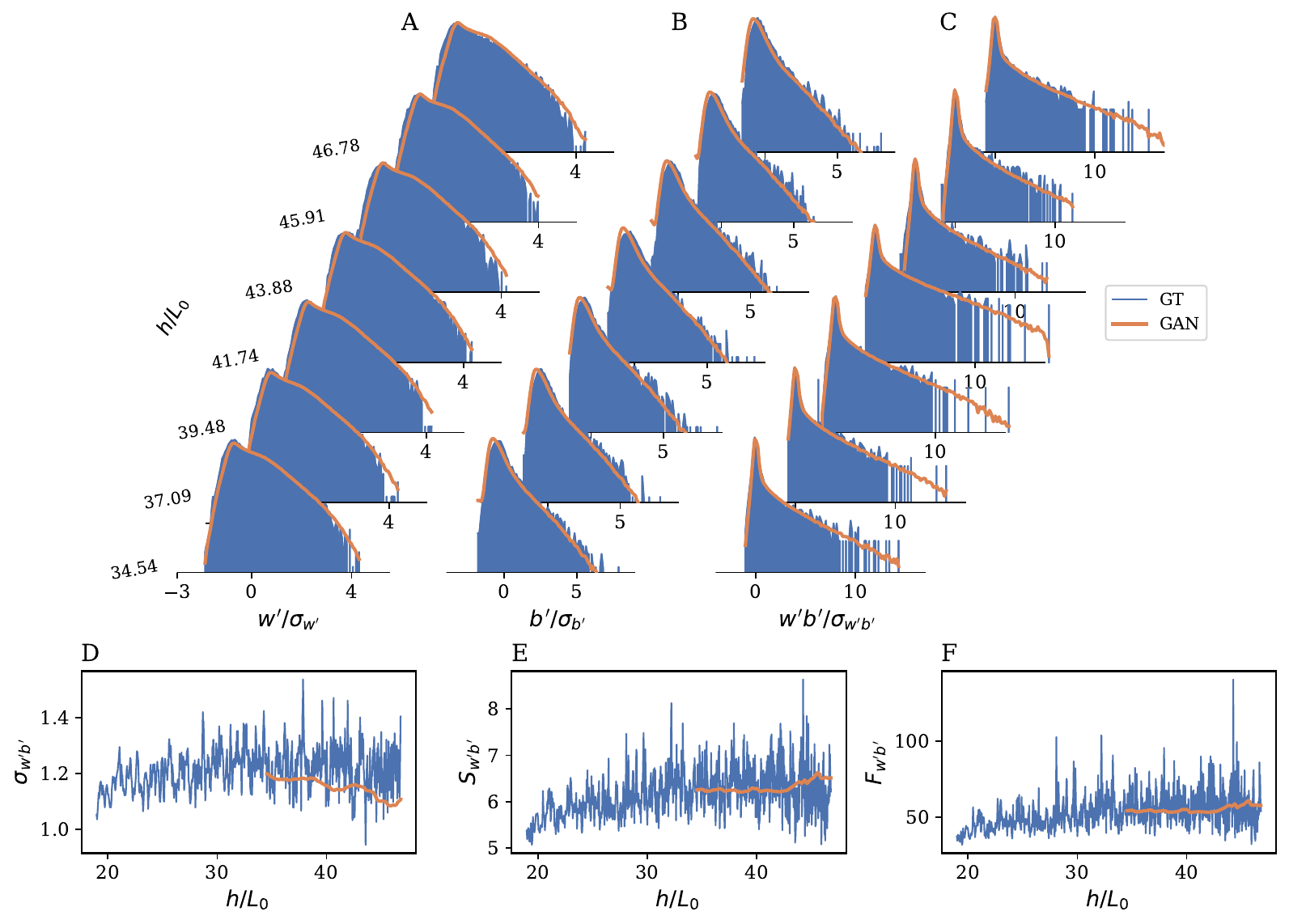}
\caption{Prediction of the time evolution of the turbulence statistics by the generative adversarial network (GAN). The probability density functions of the vertical velocity fluctuations (A), the buoyancy fluctuations (B), and the buoyancy flux (C) are shown for different times $h/L_0$. All quantities are normalized by their corresponding root mean square values. The ground truth (GT) is given in blue, the GAN result in red. The bottom row panels display the time evolution of the moments of the buoyancy flux. These are the fluctuations (D), the skewness (E), and the flatness (F). The blue line is for the GT, the orange line for the GAN. Data are for $z/h=0.5$.}
\label{fig:Fig5}
\end{figure}
%-----------------------------------------------------

\section{Generated subgrid-scale fields}
\subsection{Structure and statistics} Figure \ref{fig:Fig4}(A--F) demonstrates that the GAN can produce strikingly real horizontal slices of $w'$, $b'$, and $w'b'$ that reproduce all spatial features of all (ground truth) fields, here for $\tilde z =0.5$ at $h/L_0 = 48.41$. Moreover, the generated data reproduce the highly non-Gaussian, skewed PDFs of all three fields almost perfectly, see panels (G--I). Even far-tail events are captured fairly well in all three cases. This has direct implications for a machine learning-based convective parameterization, as information on the spatial organization of individual plumes becomes available in contrast to the physical models based on means and Gaussian statistics. Additionally, the derived PDFs allow to access the high-order statistics of the fields, such as the strongest updrafts that shoot into the stably stratified layer above the CBL. Similar results can be observed for the distinct patterns and PDFs at $\tilde z=0.2$, slightly above the surface layer, and for $\tilde z= 1.0$, in the entrainment zone, as well as at different times of the boundary layer growth, see {\color{blue} SI Appendix}. For these heights, we train two additional GANs with data from the respective horizontal planes. In this way, the ML approach enables a fast generation of spatial and statistical information at three important and representative heights, the surface and mixed layer and the entrainment zone. 

\subsection{Transient time evolution}
Figure \ref{fig:Fig5} reports the turbulence statistics with respect to time. In panels (A--C) of the figure, we show the PDFs of $w^{\prime}/\sigma_{w'}$, $b^{\prime}/\sigma_{b'}$, and $w^{\prime}b^{\prime}/\sigma_{w'b'}$ where the corresponding $\sigma^2_p(t)=\langle p^2\rangle_{x,y}$ for $p=\{w', b', w'b'\}$. It is seen that the GAN generates the correct time evolution of the statistics and the corresponding moments. In panels (D--F), we show the time evolution of the fluctuations $\sigma_{w'b'}^2=\langle(w'b')^2\rangle_{x,y}$ and normalized third- and fourth order statistical moments of the buoyancy flux versus time. The latter two are the skewness $S_{w'b'}$ and the flatness $F_{w'b'}$ which are given by 
\begin{equation}
S_{w'b'}(t)=\frac{\langle (w'b')^3\rangle_{x,y}}{\sigma_{w'b'}^3} \quad\mbox{and}\quad 
F_{w'b'}(t)=\frac{\langle (w'b')^4\rangle_{x,y}}{\sigma_{w'b'}^4}\,.
\end{equation}
The figure shows that both normalized moments deviate strongly from the Gaussian values which would follow to $S_{w'b'}=0$ and $F_{w'b'}=3$. Again, the GAN reproduces the slow growth of the moments fairly well. Note also that the fluctuations of the time series grows with time since the sampling area for taking statistics shrinks with respect to time. Flatness valus of the order of 50 indicate a highly intermittent transport across the growing CBL.

\subsection{Comparison with eddy diffusivity mass-flux} Finally, the GAN networks enable a computation of the mean turbulent buoyancy flux $\langle w'b'\rangle$ at reference heights inside the CBL. We now compare their performance with the ground truth, as well as the estimations of two frequently used closures, see \eqref{eq:mass1}. Es mentioned before, they are termed eddy diffusivity mass-flux (EDMF) and contain the mass-flux parametrization as a second term next to the Boussinesq term which always connects subgrid-scale flux (or stress) with a mean gradient via an eddy (or turbulent) viscosity. 

\begin{figure}[!htpb]
\centering
\includegraphics[width=0.5\textwidth]{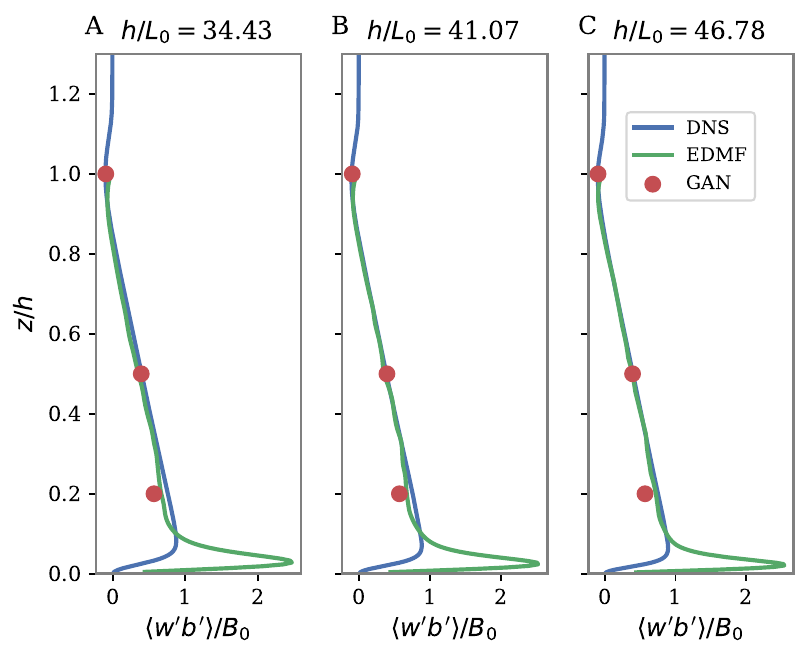}
\caption{Vertical profile of the buoyancy flux $\langle w'b'\rangle/B_0$ at three values of $h/L_0$. The eddy diffusivity mass-flux scheme uses a mass flux that is proportional to the updraft velocity $M_u \sim w_u$, which is obtained from a plume model (green curve), see also the {\color{blue}SI Appendix}. The generative network results at three heights $z/h=0.2$, $0.5$, and $1.0$ are marked as red symbols.}
\label{fig:Fig6}
\end{figure}

To this end, we applied an EDMF scheme which follows from a steady plume model \cite{Siebesma2007}. As detailed in the {\color{blue}SI Appendix}, the model comprises \eqref{eq:mass1} and the following two equations
\begin{align}
\frac{\partial \langle b\rangle_u}{\partial z} &= \epsilon (\langle b\rangle_{x,y}-\langle b\rangle_u)\,,
\label{eq:mass3a}
\\ 
\frac{1}{2}(1-2\mu) \frac{\partial \langle w\rangle_u^2}{\partial z} &+C_1 \epsilon \langle w\rangle_u^2 = \langle b\rangle_u - \langle b\rangle_{x,y}\, 
\label{eq:mass3b}
\end{align}
These equations still contain the adjustable model parameters $\mu$, $C_1$, and $\epsilon$. Typically $\mu\approx 0.15$ and $C_1\approx 0.5$ are chosen. The yet unknown entrainment rate $\epsilon$ is also height-dependent and modeled by the following expression \cite{Siebesma2007} 
\begin{align}
    \epsilon(z) \simeq 0.4\left(\frac{1}{z}+\frac{1}{h-z} \right)\,.
    \label{eq:eps}
\end{align}
Equations (\ref{eq:mass3a})--(\ref{eq:eps}) for the variables $\langle b(z)\rangle_u$ and $\langle w(z)\rangle_u$ together with \eqref{eq:mass1} are then solved numerically.
Closure EDMF takes finally $M_u \simeq a_u w_u(z)$ \cite{Soares2004} where $a_u\approx 0.05$. See again {\color{blue}SI Appendix} where we also provide the model formula for the turbulent diffusivity $\kappa_t$. The comparison of all models is shown in Fig. \ref{fig:Fig6} for three different times $h/L_0$. We find excellent agreement between the three flux values predicted by the GAN and the DNS profiles. Our ML results are also close to the profiles of the EDMF parameterization.

\section{Final discussion and outlook}
We have presented a parametrization for a dry convective boundary layer on sub-kilometer scales. Our convective parametrization is based on a generative adversarial network with Wasserstein loss metrics which provides the horizontal spatial organization and the statistics of vertical velocity component and the buoyancy, and thus the one of the local buoyancy flux at different heights inside the CBL. Our machine learning algorithm is able to predict the statistical distributions of the buoyancy flux in the course of the diurnal cycle which causes a growth of the layer height $h(t)$ and an increasing entrainment into the stable atmosphere on top of the CBL. The highly non-Gaussian and intermittent statistics of this process is successfully reproduced by our generative convective parametrization.

We have used the self-similar growth of the CBL in correspondance with the classical mixed-layer theory by Deardorff to augment the training data base for the algorithm. In this way, we incorporated  he leading-order physical effects of the boundary layer growth into the ML algorithm. This enabled us to generate synthetic snapshots of the growing CBL at three different heights which the highly asymmetric turbulence when comparing updrafts in the form of a narrow-ridge network and downdrafts in the form of broader downwellings in between. The U-net architecture of the generator is able to generate highly segmented fields with the correct spatial organisation of the convective turbulence on the basis of a relatively small number of training data. This is a significant improvement over conventional parameterization techniques like the eddy diffusivity mass-flux, which cannot provide information on the horizontal organization of strong individual plumes that overshoot in the stable layer at the top and determine the entrainment rate. Moreover, as opposed to stochastic parameterization schemes which are built on Gaussian statistics, our ML approach reproduces the non-Gaussian distributions of $w'$, $b'$, and $w'b'$. The trained generator network is generalizable to different surface buoyancy fluxes and stratification strengths of the free atmosphere, since the training is done in self-similar units of the mixed layer theory.  

Several directions for future research are possible from this point. The first is to include wind shear into the present convective boundary layer model and to extend the GAN. The cloud layer which typically start at the top of the CBL can be included by a switch from the single buoyancy field as given by \eqref{eq:buo} to a buoyancy, which is given by the liquid water potential temperature $\theta_l$, in combination with the total water content. A further direction is related to other applications of convection that include compressibility effects, e.g. for astrophysical flows where data-driven mass-flux parametrizations might substitute classical mixing length closures \cite{Kupka2017}. These studies are underway and will be reported elsewhere.     
 
%---------------------------------------------------
%-----------------------------------------------------

\section*{Acknowledgments}
F.H. is supported by the project No. P2018-02-001 "DeepTurb -- Deep Learning in and of Turbulence" of the Carl Zeiss Foundation. Partial support for J.P.M. was provided by grant PID2019-105162RB-I00 funded by MCIN/AEI/10.13039/501100011033. The authors gratefully acknowledge the Gauss Centre for Supercomputing e.V. (www.gauss-centre.eu) for funding this project by providing computing time on the GCS Supercomputer SuperMUC-NG at Leibniz Supercomputing Centre (www.lrz.de). We thank Christopher M\"uller for his initial help with the machine learning algorithm.

%----------------
% APPENDIX
%----------------
\newpage
\appendix

\section{Convective boundary layer model}
In this section, we collect the equations of motion which model the convective boundary layer together with the most important dimensionless parameters and the direct numerical simulation (DNS) method.

\subsection{Equations and dimensionless parameters}
We consider the Boussinesq limit of the Navier–Stokes equations which is given by
\begin{align}
\label{eq:app1}
\frac{\partial {\bf v}}{\partial t} + {\bf \nabla}\cdot({\bf v}\otimes {\bf v}) &= - {\bf \nabla}\pi + \nu {\bf \nabla}^2{\bf v} + b{\bf e}_z\,,\\ 
\label{eq:app2}
{\bf\nabla}\cdot {\bf v} & = 0 \,,\\ 
\label{eq:app3}
\frac{\partial b}{\partial t} + {\bf \nabla}\cdot({\bf v}b) & = \kappa {\bf \nabla}^2 b\,,
\end{align}
where $\pi$ is a modified dynamic pressure or kinematic pressure, i.e., a deviation from the pressure in the hydrostatic balance divided by a reference density. Quantity $\nu$ is the kinematic viscosity, $\kappa$ is the buoyancy diffusivity. 

The velocity field ${\bf v}({\bf x},t)=(u,v,w)$ satisfies no-slip boundary conditions at the bottom and free-slip boundary conditions at the top \cite{Garcia2014}. The buoyancy field $b({\bf x},t)$ satisfies Neumann boundary conditions at the bottom and the top. The surface buoyancy flux is given by $B_0$. Periodic boundary conditions are imposed in the horizontal directions for all fields. The initial condition is ${\bf v} = 0$ and, for the buoyancy,
\begin{equation}
b_{\rm bg}(z)=N_0^2 z+f(z) \,,
\label{eq:bg}
\end{equation}
where $f(z)$ is the variation near the bottom surface to satisfy $-\kappa \partial b/\partial z = B_0$. The function $f(z)$ is only non-zero very near the surface and we consider long enough times for the effects of $f(z)$ on the results to be negligibly small. The quantity $N_0$ is the Brunt-V\"ais\"al\"a frequency in \eqref{eq:bg}. The Prandtl number $Pr$ and the buoyancy Reynolds number are the two central dimensionless parameters in this setup. They are given by 
\begin{equation}
Pr=\frac{\nu}{\kappa}\,,\quad\quad Re_0=\frac{B_0}{\nu N_0^2}\,.
\label{eq:param}
\end{equation}
Times are expressed in units of $N_0^{-1}$ and lengths in units of the Ozmidov-type scale which is given by 
\begin{equation}
L_0=\left(\frac{B_0}{N_0^3}\right)^{1/2}\,.
\label{eq:param1}
\end{equation}
The Prandtl number is $Pr=1$ in the present DNS. %direct numerical simulations (DNS). 

\subsection{Convective boundary layer and encroachment height}
In the convective boundary layer, there is no upper lid that confines the motion; a turbulent boundary layer forms next to the bottom surface. The difference to standard Rayleigh-B\'{e}nard convection is that here the boundary layer is only caused by turbulent free convection, forced by a constant heat flux at the bottom surface. The convective boundary layer (CBL) warms and thickens with respect to time. The top boundary of the computational domain is placed far enough to not influence the boundary-layer properties, and otherwise the height of the computational domain becomes unimportant for the analysis.

Except for the CBL, the density in the region above the boundary layer is stably stratified, which hinders the growth of the boundary layer. This region above the boundary layer is referred to as {\em free troposphere} or {\em free atmosphere}; it is equivalent to the free-stream in a shear-driven boundary layer. One effect of this stratification is that the transition region from the turbulent boundary layer to the non-turbulent free stream is thinner than in the case without stratification. This is because the stratification of the free troposphere hinders the vertical displacement of fluid particles. This transition region is referred to as the {\em entrainment zone}. Another effect is that this stratification hinders the growth of the CBL thickness, to the point that the characteristic time for changes of mean properties is small compared to the large eddy turnover time, and a statistically quasi-steady state is observed.

This stratification of the free-stream region is imposed by setting the initial condition of the buoyancy field as $b_\mathrm{bg}=N_0^2z$, see \eqref{eq:bg}. The transition region between the turbulent boundary layer and the free tropopause (i.e., the entrainment zone) occurs where the mean profile of buoyancy, $\langle b(z)\rangle$, changes from the vertically constant value inside the turbulent CBL to the linear profile $b_\mathrm{bg}=N_0^2z$ in the non-turbulent free troposphere, see also Fig. 1(G) in the main text. We refer to the vertical location of this thin transition region as CBL top. There are various ways to define the CBL top and thus the thickness of the whole convective boundary layer. To a first approximation, they all are proportional to the {\em encroachment height} which is defined as \cite{Carson:1975,Garcia2014}
\begin{equation}
    h(t)= \left\{2N_0^{-2}\int_0^{z_\infty}(\langle b(z)\rangle_{x,y} - b_\mathrm{bg}(z))\mathrm{d}z\right\}^{1/2} \,.
    \label{equ:zenc}
\end{equation}
From the point of view of dimensional analysis, we have substituted the cell height in a Rayleigh-B\'{e}nard setup by $L_0$ which was defined in  \eqref{eq:param1}. It characterizes the thickness of the transition region at the top of the CBL. The reference length $L_0$ is constant, but $h$ increases with time $t$. Consider the mean buoyancy equation with statistical homogeneity in the horizontal directions. Then follows from \eqref{eq:app3}
\begin{equation}
\frac{\partial}{\partial t}[\langle b\rangle_{x,y}-b_{\rm bg}(z)] =-\frac{\partial}{\partial z}\left[\langle w^{\prime}b^{\prime}\rangle_{x,y} -\kappa \frac{\partial \langle b\rangle_{x.y}}{\partial z}\right]\,,
\label{equ:zenc1}
\end{equation}
where we have added the time-independent profile $b_{\rm bg}(z)$ on the left hand side. Integrate \eqref{equ:zenc1} from $z=0$ to $z=z_{\infty}$ and use the fact that far away from the surface $\langle b\rangle_{x,y}\to b_{\rm bg}(z)=N_0^2z$. This results to
\begin{equation}
\frac{\partial}{\partial t}\int_0^{z_{\infty}}[\langle b\rangle_{x,y}-b_{\rm bg}(z)] dz =\kappa N_0^2 +B_0=B_0\left(1+\frac{1}{Pr Re_0}\right)\,.
\label{equ:zenc2}
\end{equation}
Thus an additional integration with respect to time, together with \eqref{equ:zenc}, gives
\begin{equation}
\int_0^{z_{\infty}}[\langle b\rangle_{x,y}-b_{\rm bg}(z)] dz =B_0\left(1+\frac{1}{Pr Re_0}\right)(t-t_0)=\frac{h^2N_0^2}{2}\,.
\label{equ:zenc3}
\end{equation}
Encroachment height and Ozmidov-type scale, see \eqref{eq:param1}, are then related according to
\begin{equation}\label{equ:zencl0}
    \frac{h(t)}{L_0}=\left[2N_0(t-t_0)\left(1+\frac{1}{Pr Re_0}\right)\right]^{1/2} \;,
\end{equation}
where $t_{0}$ is an integration constant. See also \eqref{eq:app3} in the main text for $Pr=1$. The growth rate of $h(t)$ is referred to as mean entrainment velocity
\begin{equation}\label{equ:we}
    w_\mathrm{e}\equiv\frac{\mathrm{dh}}{\mathrm{d}t}\,.
\end{equation}
The convective Rayleigh number, which is given by 
\begin{equation}
    Ra_c(t)=\frac{B_0h(t)^4}{\nu\kappa^2}\,,
\end{equation}
increases as $h(t)$ increases with time $t$.

\subsection{Direct numerical simulations}
Discretization in space is performed using sixth-order spectral-like compact finite differences on a structured Cartesian grid \cite{Mellado2012}. A low-storage fourth-order accurate Runge–Kutta scheme is used for time stepping. The discrete solenoidal constraint is satisfied to machine accuracy using a Fourier decomposition along the periodic horizontal planes in $x$-$y$ directions and a factorization of the resulting set of equations along the vertical coordinate $z$. The local ratio between the vertical grid spacing, $\Delta z$, and the Kolmogorov scale, $\eta_K$, is $\Delta z/\eta_K\lesssim 1.2$.

\section{Mass-flux parametrization}
In this section, we describe in brief the basics of the mass-flux parametrization which is used in many global numerical models of the Earth system, such as the Integrated Forecasting System of the European Centre for Medium-Range Weather Forecasts (\url{www.ecmwf.int/en/publications/ifs-documentation}). The mass-flux parametrization is one part of the eddy diffusivity mass-flux which models the vertical profiles of the subgrid-scale fluxes of liquid-water potential temperature $\theta_l$ (convective heat flux) or total water content $q_t$ (moisture flux) across the atmosphere. 

\subsection{Buoyancy flux closure by eddy diffusivity mass-flux}
In the present proof-of-concept study we omit phase changes and cloud formation. The liquid-water potential temperature $\theta_l$ reduces then to the virtual potential temperature $\theta$ and the buoyancy $b$ becomes
\begin{equation}
b({\bf x},t)=g\frac{\theta_{\rm v}({\bf x},t)-\theta_{{\rm v},0}}{\theta_{{\rm v},0}}\,,
\end{equation}
with the acceleration due to gravity $g$ and the virtual potential temperature at the surface $\theta_{{\rm v},0}$. The eddy diffusivity mass-flux closure for the sub-grid scale buoyancy flux is given by \cite{Siebesma1996,Siebesma2007,Witte2022},
\begin{equation}
\rho_0\langle w^{\prime}b^{\prime}(z)\rangle_{x,y} = -\rho_0\kappa_t \frac{\partial \langle b\rangle_{x,y}}{\partial z} + M_u (\langle b(z)\rangle_u-\langle b(z)\rangle_{x,y})\,.
\label{eq:app_mass1}
\end{equation}
Here, $\rho_0$ is the constant reference density in the lower atmospheric boundary layer and $M_u$ is the unknown mass flux. As already said in the main text, the horizontal cross section $A$ can be decomposed into disjoint updraft regions (u) and the remaining environment (e), i.e., $A_{\rm total}=A_u \cup A_e$. For simplicity we will denote $A_{\rm total}$ simply by $A$ to ease notation. The first term in \eqref{eq:app_mass1} is the standard Boussinesq term for a turbulent stress which contains the eddy diffusivity $\kappa_t$ and the mean buoyancy gradient. The additional second term comprises the mass-flux parametrization. In detail, 
\begin{align}
\langle w^{\prime}b^{\prime}(z)\rangle_{x,y}& =\frac{1}{A}\int_A (w-\langle w\rangle_{x,y})(b-\langle b\rangle_{x,y}) \,dA\,,\label{eq:app_mass2a}\\
a_u\langle w^{\prime}b^{\prime}(z)\rangle_{u}& =\frac{1}{A}\int_A (w-\langle w\rangle_{u})(b-\langle b\rangle_{u}) \,dA\,,\label{eq:app_mass2b}\\
(1-a_u)\langle w^{\prime}b^{\prime}(z)\rangle_{e}& =\frac{1}{A}\int_A (w-\langle w\rangle_e)(b-\langle b\rangle_{e}) \,dA\,,\label{eq:app_mass2c}
\end{align}
with the area fractions $a_u=A_u/A$ and $1-a_u=A_e/A$. We neglect the first gradient term on the right hand side of \eqref{eq:app_mass1} for a while. The second term of \eqref{eq:app_mass1} can be rewritten to 
\begin{equation}
\langle w^{\prime}b^{\prime}\rangle_{x,y} = a_u \langle w^{\prime}b^{\prime}\rangle_{u} + (1-a_u)\langle w^{\prime}b^{\prime}\rangle_{e} + a_u(1-a_u) [\langle w\rangle_u-\langle w\rangle_{e}][\langle b\rangle_u-\langle b\rangle_{e}]\,.
\label{eq:app_mass3}
\end{equation}
The first term quantified the buoyancy flux connected with $A_u$, the second the one with $A_e$, and the third is an exchange term due to organized turbulence which connects both. The following approximations can be made: (i) $a_u\ll 1$, in practice mostly about 5 \%; (ii) $\langle w\rangle_e\simeq \langle w\rangle_{x,y}$, and (iii) $\langle b\rangle_e\simeq \langle b\rangle_{x,y}$. This results to the mass-flux parametrization
\begin{equation}
\langle w^{\prime}b^{\prime}(z)\rangle_{x,y} \simeq \frac{M_u}{\rho_0} (\langle b(z)\rangle_u-\langle b(z)\rangle_{x,y})\quad \mbox{with}\quad M_u=\rho_0 a_u [\langle w(z)\rangle_u-\langle w(z)\rangle_{x,y}] \,,
\label{eq:app_mass4}
\end{equation}
with the undetermined mass flux $M_u$. Recall also that the mass flux carries the physical dimension of kg$\,$m$^{-2}$s$^{-1}$. A plume model is required to determine this unknown quantity which is discussed in the following.  

\subsection{Steady plume model} \eqref{eq:app3} of the present model is the counterpart of the set of prognostic equations in a global circulation model. We apply a Reynolds decomposition and obtain the following unclosed coarse-grid equation \cite{Siebesma1995}
\begin{equation}
\frac{\partial \langle b(z)\rangle_{x,y}}{\partial t} = -\frac{1}{\rho_0} \frac{\partial}{\partial z} [\rho_0 \langle w^{\prime} b^{\prime}(z)\rangle_{x,y}] + \frac{\partial \langle b(z)\rangle_{x,y}}{\partial t} \Bigg|_{\rm forcing}\,.
\label{eq:app_plu1}
\end{equation}
All remaining terms due to horizontal mean advection and diffusion are summarized in the last term of \eqref{eq:app_plu1} as a forcing term. Note that we get in this way  an effective {\em vertical} transport model without any information on the spatial horizontal mesoscale organisation of the convective turbulence which the DNS in the main text display. The area decomposition into $A_u$ and $A_e$ gives 
\begin{equation}
\langle b(z)\rangle_{x,y} = a_u \langle b(z)\rangle_{u} + (1-a_u) \langle b(z)\rangle_{e}\,.
\label{eq:app_plu2}
\end{equation}
Equations \eqref{eq:app_mass3} and \eqref{eq:app_plu2} are now plugged into \eqref{eq:app_plu1}. We use that $a_u(1-a_u)\approx a_u$ since $a_u\ll 1$ and apply additionally the approximations (ii) and (iii) from the last subsection. This leads to the following updraft budget equation (omitting the $z$ argument),
\begin{align}
\rho_0\frac{\partial}{\partial t} [a_u\langle b\rangle_{u}] = 
-\frac{\partial}{\partial z} [\rho_0 a_u \langle w^{\prime} b^{\prime}\rangle_{u}] -\frac{\partial}{\partial z} [\underbrace{\rho_0 a_u (\langle w\rangle_{u}-\langle w\rangle_{x,y})}_{=M_u}\langle b\rangle_u] + E\langle b\rangle_e -D\langle b\rangle_u
+\rho_0a_u\frac{\partial\langle b\rangle_{x,y}}{\partial t} \Bigg|_{\rm forcing}\,,
\label{eq:app_plu3a}
\end{align}
and the corresponding downdraft budget equation
\begin{align}
\rho_0\frac{\partial}{\partial t} [(1-a_u)\langle b\rangle_{e}] = 
-\frac{\partial}{\partial z} [\rho_0 (1-a_u) \langle w^{\prime} b^{\prime}\rangle_{e}] +\frac{\partial}{\partial z} [M_u\langle b\rangle_e] - E\langle b\rangle_e +D\langle b\rangle_u
+\rho_0(1-a_u)\frac{\partial\langle b\rangle_{x,y}}{\partial t} \Bigg|_{\rm forcing}\,.
\label{eq:app_plu3b}
\end{align}
We have added two terms in both equations to express the mass exchange due to entrainment and detrainment with rates $E$ and $D$, respectively. Adding both equations brings us back to \eqref{eq:app_plu1}. Furthermore, a continuity equation is required which can be obtained from one of the budget equations by substituting $\langle b\rangle$ with 1. This gives
\begin{equation}
\rho_0\frac{\partial a_u}{\partial t} =-\frac{\partial M_u}{\partial z} +E -D \,.
\label{eq:app_plu3c}
\end{equation}
Next a (statistically) steady state is assumed such that \eqref{eq:app_plu3a} and \eqref{eq:app_plu3c} simplify to, see also ref. \cite{Siebesma2007},
\begin{align}
\frac{\partial}{\partial z} [M_u\langle b\rangle_u] &= 
-\frac{\partial}{\partial z} [\rho_0 a_u \langle w^{\prime} b^{\prime}\rangle_{u}] + M_u\epsilon\langle b\rangle_e -M_u\delta\langle b\rangle_u\,,
\label{eq:app_plu4a}\\
\frac{\partial M_u}{\partial z} &= M_u (\epsilon-\delta)\,,
\label{eq:app_plu4b}
\end{align}
where fractional entrainment and detrainment rates follow from $E=M_u\epsilon$ and $D=M_u\delta$, respectively. When the first term on the right hand side of \eqref{eq:app_plu4a} is set to zero, both equations can be combined to give the following simple updraft equation
\begin{equation}
\frac{\partial \langle b\rangle_u}{\partial z} = \epsilon (\langle b\rangle_{x,y}-\langle b\rangle_u)\,. 
\label{eq:app_plu6}
\end{equation}
Furthermore, we need an equation for $\langle w\rangle_u$ such that we can close the buoyancy flux model \cite{Schumann1991,Siebesma2007}. This is obtained from the momentum balance for the updrafts, assuming again a steady regime. This gives 
\begin{equation}
-\frac{1}{2}\frac{\partial \langle w\rangle_u^2}{\partial z} - C_1 \epsilon \langle w\rangle_u^2 - \frac{1}{\rho_0}\frac{\partial \langle p\rangle_{x,y}}{\partial z}  + \langle b\rangle_u-\langle b\rangle_{x,y} =0 \,. 
\label{eq:app_plu7}
\end{equation}
This equation went again through a number of simplification steps: (i) the first term results from the nonlinear advection of $\langle w\rangle_u$, (ii) the second term encloses all remaining advection terms, (iii) the effective entrainment rate which is connected with the turbulent mixing in (ii) is assumed to be directly proportional to the entrainment for $\langle b\rangle_u$ in \eqref{eq:app_plu6}. (iv) The pressure gradient term is approximated finally to be $\partial (\mu \langle w\rangle_u)/\partial z$. This is the last step and the {\em mass-flux parametrization} comprises the following coupled system of equations which would have to be solved together with the mean prognostic equation for $\langle b\rangle_{x,y}$,
\begin{align}
\frac{\partial \langle b\rangle_u}{\partial z} &= \epsilon (\langle b\rangle_{x,y}-\langle b\rangle_u)\,,
\label{eq:app_mf1a}\\
\frac{1}{2}(1-2\mu)\frac{\partial \langle w\rangle_u^2}{\partial z} + C_1 \epsilon \langle w\rangle_u^2 &= \langle b\rangle_u-\langle b\rangle_{x,y} \,.
\label{eq:app_mf1b}
\end{align}
If we know $\langle w\rangle_u$ and $\langle b\rangle_u$ then we can calculate the second term on the right hand side of the EDMF model in \eqref{eq:app_mass1} for a given $a_u$. The model parameters $\epsilon$, $C_1$, and $\mu$ are however unknown. Large eddy simulations studies by Siebesma et al. \cite{Siebesma2007} suggest $C_1\simeq 0.5$ and $\mu\simeq 0.15$. The fractional entrainment rate $\epsilon$, which will be height dependent, is more recently included as a stochastic multi-plume model, i.e., for each plume $i$ the EDMF model equations are solved. The entrainment rates $\epsilon_i$ are chosen from a Poisson distribution \cite{Witte2022}.    

\subsection{Numerical Implementation}
\begin{figure}[!htbp]
\centering
\includegraphics[width=.6\textwidth]{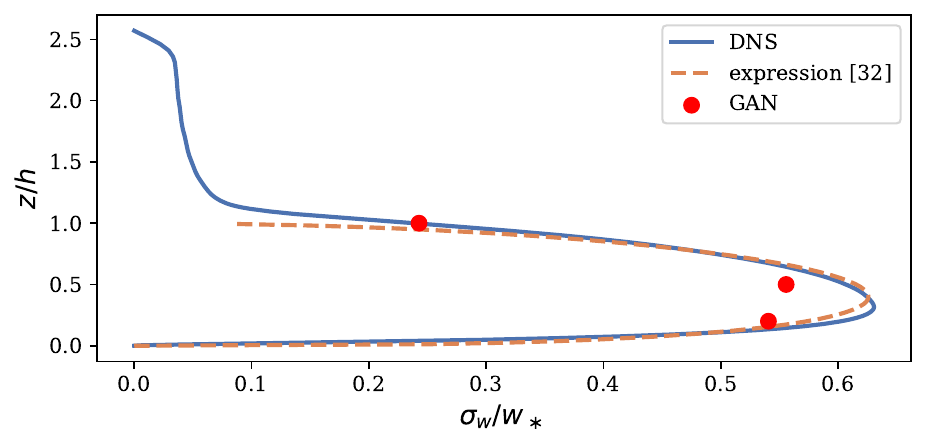}
\caption{Standard deviation of vertical velocity $w$ obtained from the direct numerical simulation (blue) in comparison to the expression \eqref{eq:app_sigma_w} (orange). The GAN values are shown in red.}
\label{fig:Appendix_Fig1}
\end{figure}

The eddy-diffusivity $\kappa_t$ is implemented by a simple profile method  \cite{Holtslag1998} and results to
\begin{align}
    \kappa_t(z) = k\left(39k\frac{z}{h}\right)^{1/3}\frac{z}{h}\left(1-\frac{z}{h}\right)^2  w_\ast h\,,
\end{align}
where $k\approx 0.4$ is the von K\'arm\'an constant. The mass-flux contributions were obtained by integrating equations \eqref{eq:app_mf1a} and \eqref{eq:app_mf1b} using a fourth-order Runge-Kutta scheme. We choose the starting height $z_0>0$ together with the boundary conditions
\begin{align}
    \langle b(z_0)\rangle_u &= \langle b(z_0)\rangle + \alpha B_0/\sigma_w,\\
    \langle w(z_0)\rangle_u^2 &= \sigma_w^2(z_0),
\end{align}
where we make use of an empirical standard deviation $\sigma_w$ as in ref.~\cite{Holtslag1991} which is given by 
\begin{align}
\label{eq:app_sigma_w}
    \sigma_w(z) \simeq 1.3\left(0.6\frac{z}{h}\right)^{1/3} \left( 1-\frac{z}{h}\right)^{1/2} w_\ast,
\end{align}
which fits well for the present direct numerical simulations as seen in Fig. \ref{fig:Appendix_Fig1}.
%----------------------------------------
%----------------------------------------
Furthermore, we use an entrainment rate profile as suggested in \cite{vanUlden1997, Siebesma2007} which is given by
\begin{align}
    \epsilon \simeq 0.4\left(\frac{1}{z}+\frac{1}{h-z} \right).
\end{align}
These input lead to the profiles in Fig. 5 of the main text which we used to compare this closure with our machine learning results.

\section{Generative Adversarial Network}
In this section, we describe the machine learning algorithm which is used to generate synthetic data. We explain the architecture of the network, details on the Wasserstein loss metrics, and our augmentation of the training data base.  

\subsection{Generator and discriminator architecture}
A Generative Adversarial Network (GAN)~\cite{Goodfellow2014} is a neural network comprising two interconnected and adversarial components: a generator and a discriminator (or critic). The generator generates synthetic data, in our case, horizontal slices of an atmospheric boundary layer simulation. The discriminator's role is to distinguish between the synthetic data produced by the generator and real data from the DNS. The two components are trained simultaneously in a competitive manner, with the generator striving to generate data that can fool the critic and the critic continuously improving its ability to classify the data correctly. This adversarial relationship drives the GAN to produce increasingly realistic, high-quality synthetic data. The training of the network is completed once the {\em Nash equilibrium} is reached \cite{Nash1950}; the discriminator is unable to distinguish synthetically generated and ground truth data. Then the generator can be decoupled from the discriminator and operate autonomously. 

Specifically, we use the Wasserstein--GAN (WGAN) formulation from Arjovsky \textit{et al.}~\cite{Arjovsky2017}, which solves the vanishing gradient problem of the generator of the original GAN model. This is done using the Wasserstein distance (or Earth-movers distance) between the synthetic and real data in the WGAN loss function. In this way, the generator network will overcome the initial phase, where the discriminator will quickly identify the generated data samples and the GAN training ends in favor of the critic network. Furthermore, the WGAN shows greater stability and convergence during the training.

Here, we use a generator with a U-shaped deep neural network structure, in short U-Net~\cite{Ronneberger2015,Fonda2019}. It has seven contraction and seven subsequent expansion layers. After each contraction layer, the spatial dimensions are halved via a convolution operation, while the channel dimensions are doubled. The opposite is done in the expansion path via transposed convolutions. Both paths are connected via residual connections, allowing for a direct information flow between two layers in the contraction and expansion paths of the U-shape. The activation function was chosen as the \textit{Parametric Rectified Linear Unit}~\cite{He2015} (PReLU)
\begin{align}
    {\rm PReLU}(x) = \left\{
    \begin{array}{lr}
        x, & \text{if } x \ge 0\\
        ax, & {\rm otherwise}
        \end{array}
        \right.
\end{align}
where the slope $a$ is a learnable parameter. The discriminator network consists of eight convolution layers, which reduce the input data to a scalar score. Here, the PReLU activation was used as well. Finally, to increase the variance of the network, we use Dropout layers, which set the layer output to zero with a probability of $0.3$.

\subsection{Loss metrics}
In the WGAN formulation, the output of the critic corresponds to a score that grades the origin of the input. Therefore, the critic is trained to output a higher score for the DNS input $\mathbf{x}$ than for its synthetic counterpart $\mathbf{x}_G$. The generator on the other hand, is trained to maximize the critic's output for $\mathbf{x}_G$. The Wasserstein distance is incorporated into the discriminator loss function using the Kantorovich-Rubinstein duality~\cite{Villani2008}. However, this restricts the critic $D(\cdot)$ to be a 1-Lipschitz function (i.e., the Lipschitz constant is one), which can be enforced by an additional gradient penalty term~\cite{Gulrajani2017}. The WGAN discriminator loss function then reads
\begin{align}
    \label{eq:ML1}  
    \mathcal{L}_{\rm D} = \underset{\mathbf{x}_G \sim P_{\rm G}}{\mathbb{E}}\left[D(\mathbf{x}_G)\right]  - \underset{\mathbf{x} \sim P_{\rm r}}{\mathbb{E}}\left[D(\mathbf{x})\right] + \lambda \underset{\hat{\mathbf{x}} \sim P_{\hat{\mathbf{x}}} }{\mathbb{E}} \left[\left(\|\nabla_{\hat{\mathbf{x}}} D\left(\hat{\mathbf{x}}\right)\|_2 - 1\right)^2\right],
\end{align}
where $\mathbf{x}$ is sampled from the real, i.e., DNS data distribution $P_{\rm r}$ and $\mathbf{x}_G$ from the generator model distribution $P_{\rm G}$. The latter is given by $G\left(\mathbf{\xi}\right)$ with the random latent variable $\mathbf{\xi} \sim \mathcal{N}(0,1)$. The last term in \eqref{eq:ML1} is the gradient penalty term, which enforces the 1-Lipschitz condition on the discriminator. The random samples $\hat{\mathbf{x}}$ are sampled from straight lines $\chi \mathbf{x} + (1-\chi)\mathbf{x}_G$ between pairs of $\mathbf{x}$ and $\mathbf{x}_G$, as proposed in by Gulrajani \textit{et al.}~\cite{Gulrajani2017}. Here, $\chi \sim U[0,1]$, i.e., uniformly distributed between 0 and 1. The gradient penalty weight $\lambda$ is a hyperparameter. Moreover, the generator loss is given by 
\begin{align}
    \mathcal{L}_{\rm G} = -\underset{\mathbf{x}_G\sim P_{\rm G}}{\mathbb{E}}[D(\mathbf{x}_G)].
    \label{eq:ML2}  
\end{align}
Finally, the Wasserstein loss function, which is given by
\begin{align}
    \mathcal{L}_{\rm WGAN} = \underset{\mathbf{x}\sim P_{\rm r}}{\mathbb{E}}[D(\mathbf{x})] - \underset{\mathbf{x}_G\sim P_{\rm G}}{\mathbb{E}}[D(\mathbf{x}_G)]\,,
    \label{eq:ML3}  
\end{align}
reflects the WGAN performance. We summarize the WGAN training procedure in Algorithm 1.
%----------------------------------------
\begin{figure}
	\includegraphics[width=\textwidth]{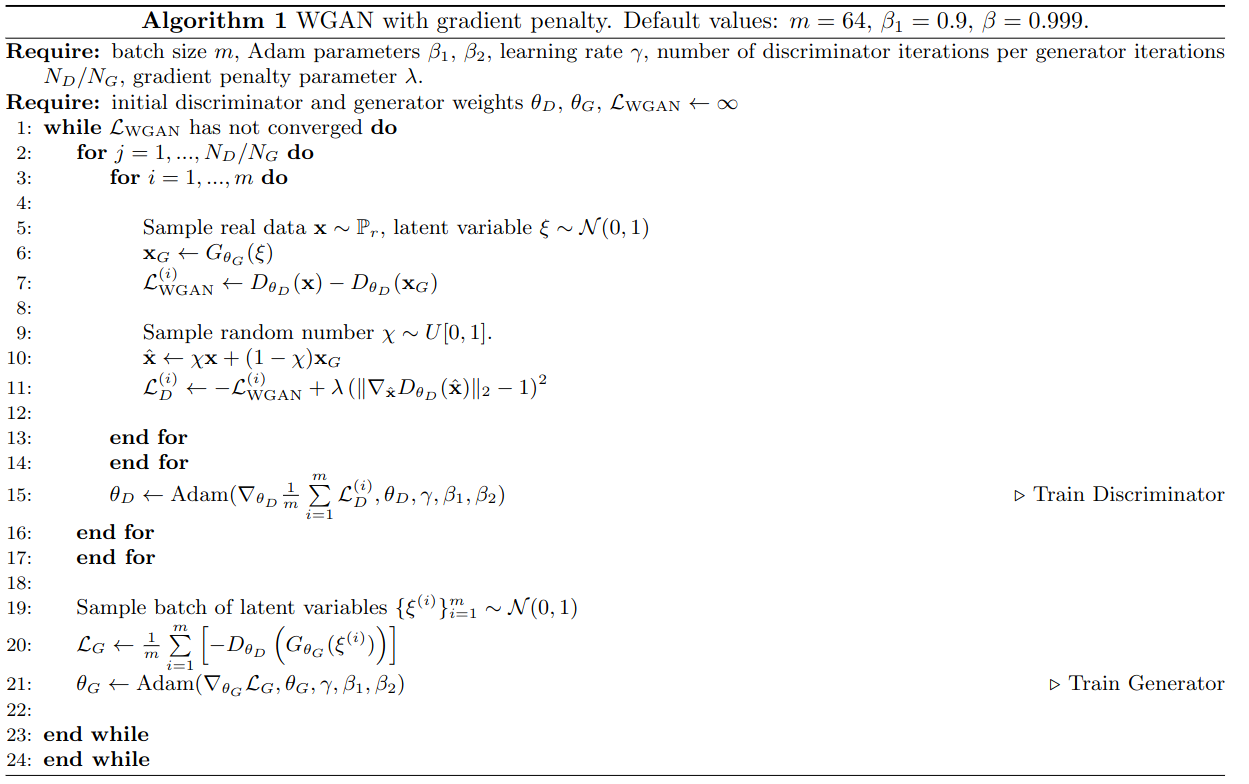}
	\caption*{}
\end{figure}
%----------------------------------------

\subsection{Physics-informed augmentation of training data base}
\begin{figure}[!htbp]
\includegraphics[width=\textwidth]{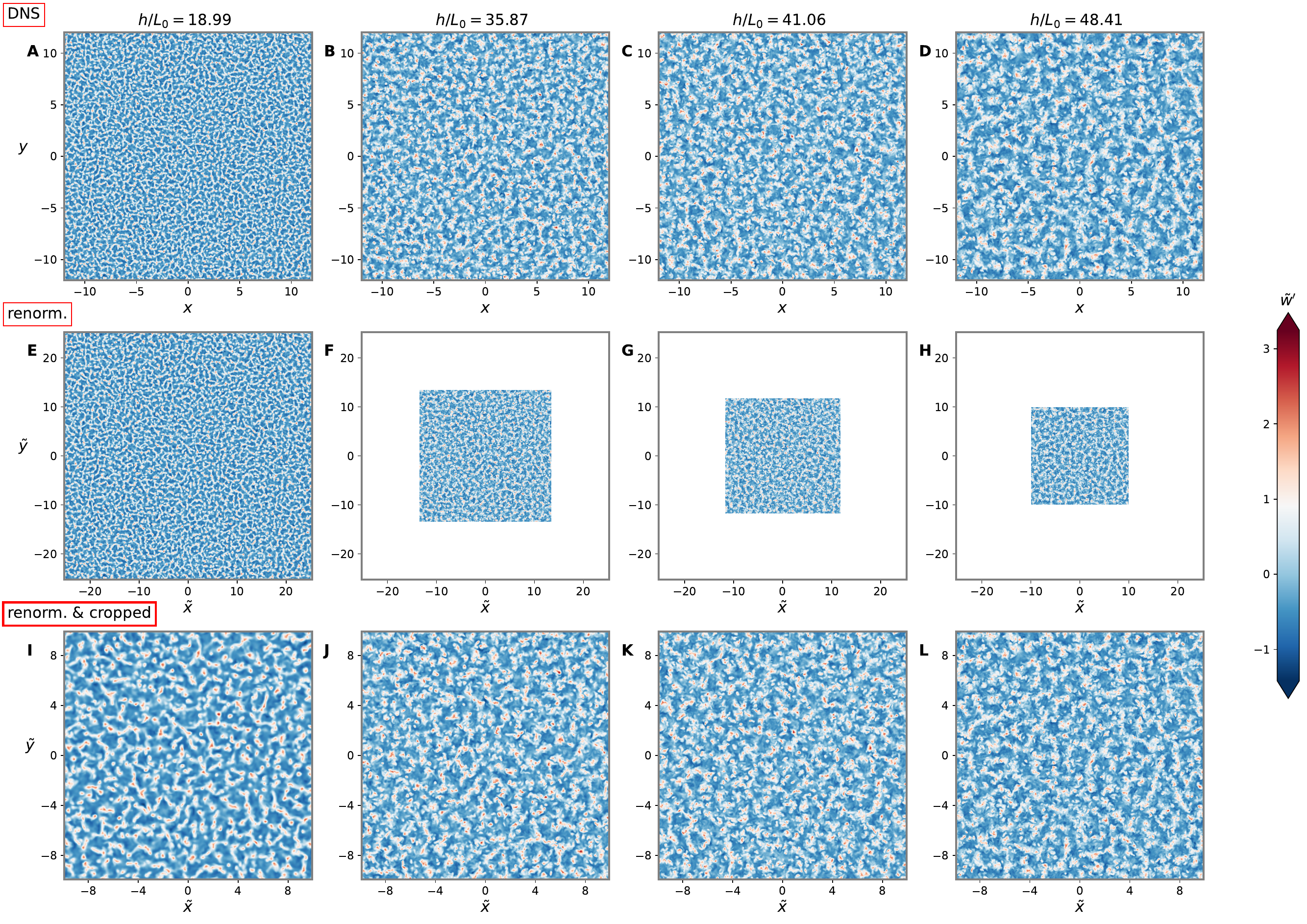}
\caption{Renormalization and cropping procedure illustrated for the vertical velocity fluctuations $\tilde{w}^{\prime}$ in plane $\tilde{z}=0.5$. (A--D) Original simulation snapshots of $w^{\prime}(x,y)$. (E--H) Rescaled fluctuations $\tilde{w}^{\prime}(\tilde{x},\tilde{y})$ in the new coordinate system $(\tilde{x},\tilde{y},\tilde{z})$. (I--L)  $\tilde{w}^{\prime}(\tilde{x},\tilde{y})$ cropped to the domain $\tilde{x}, \tilde{y} \in [-\tilde L,\tilde L]$ with $\tilde L\approx 8.5$ at time $h/L_0 = 48.41$ of panel (L). The data preparation for the buoyancy fluctuations proceeds along the same lines.}
\label{fig:Appendix_Fig2}
\end{figure}

As discussed in the main text, we renormalize the DNS data to leverage the self-similar behavior of the CBL. For this, we renormalize the fluctuations of vertical velocity and buoyancy fields by their corresponding convective scales~\cite{Deardorff:1970}, see eqns. (4) in the main text, where the CBL height $h(t)$ is given by definition (1) in the main text. Further, the coordinates are measured in units of $h(t)$, i.e., $\tilde{x} = x/h$, $\tilde{y}=y/h$, $\tilde{z} = z/h$. By doing so, we fix the CBL height. This has implications for the horizontal extent of the normalized snapshots as the aspect ratio of the domain decreases over time. Hence, to compare two snapshots at two different $h/L_0$ with the same horizontal extend, one has to crop the snapshot with the smaller value of $h(t)/L_0$. In this way, we remove the first-order effects of the transient CBL growth in a fixed plane $\tilde{z} = {\rm const}$. This process is illustrated in Fig. \ref{fig:Appendix_Fig2}.
%----------------------------------------
%----------------------------------------
In panels (I--L), it can be seen that for sufficiently large values of $h/L_0$, this procedure produces horizontal slices with comparable flow patterns. The same does hold for the buoyancy fluctuations $b'$ (not shown). It is these flow patterns that the GAN will be trained with to generate synthetic data.

\begin{figure}[!htbp]
 \centering
\includegraphics[width=\textwidth]{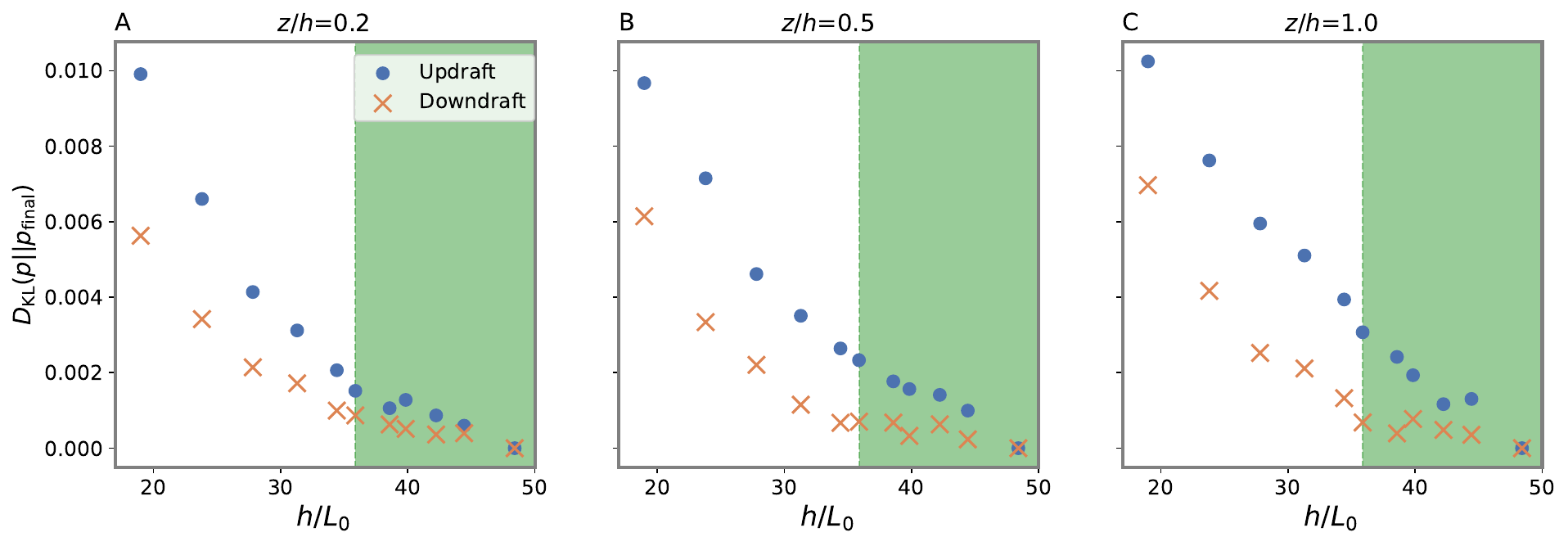}
\caption{Kullback-Leibler divergence between the probability density functions of normalized updraft (downdraft) areas $A_u$ ($A_d$) at $h/L_0$ and the corresponding final probability density function. The green shaded area marks the snapshots that were considered for the training of the GAN.}
\label{fig:Appendix_Fig3}
\end{figure}

The cutoff value of $h/L_0$, after which all re-normalized snapshots can be considered for the training routine, is determined by the following procedure. First, the individual up- and downdraft areas $A_u$ and $A_d$, given by values $w>0.95 \max(w)$ and $w<0.95\min(w)$ respectively, are retrieved using the algorithm introduced in \cite{Suzuki1985} via the \textit{OpenCV}~\cite{Itseez2015} library implementation. Finally, we computed the \textit{Kullback-Leibler divergence} (KLD) between the PDF $p$ of up- or downdrafts at $h/L_0$ with the final PDF $p_{\rm final}$ at $h/L_0=48.41$. For the updrafts, it is then given by
\begin{align}
    D_{\mathrm{KL}}(p\|p_{\rm final}) = \sum_{A_u} p(A_u) \log \left(\frac{p(A_u)}{p_{\rm final}(A_u)}\right)\,.
\end{align}
Figure \ref{fig:Appendix_Fig3} shows the KLD for the three planes over $h/L_0$. The slope of the KLD becomes less steep near $h/L_0 = 35.87$ (start of green shaded area). After this value, we consider all snapshots for the GAN training.
%----------------------------------------
%----------------------------------------

\subsection{Network training}
The normalized and cropped horizontal slices at the three heights are subsequently used to train three independent generative networks. For each height, $550$ snapshots are available for training. Moreover, we use data augmentation to increase the number of training samples to $32 \cdot 550 = 17600$. For this, we consider a quarter of the renormalized DNS data as discriminator input. Hence, also the generator will be trained to generate a quarter of the original horizontal extent. Furthermore, we add rotations by angles of $\pi/2,\pi,3\pi/2$, reflections about the $\tilde{x}$- and $\tilde{y}$-axes, as well as combinations of $\pi/2$-rotations and $\tilde{x}/\tilde{y}$-reflections. This significantly increased the number of training samples, which resulted in better generator performances. 

\begin{figure}[!htbp]
\centering
\includegraphics[width=.8\textwidth]{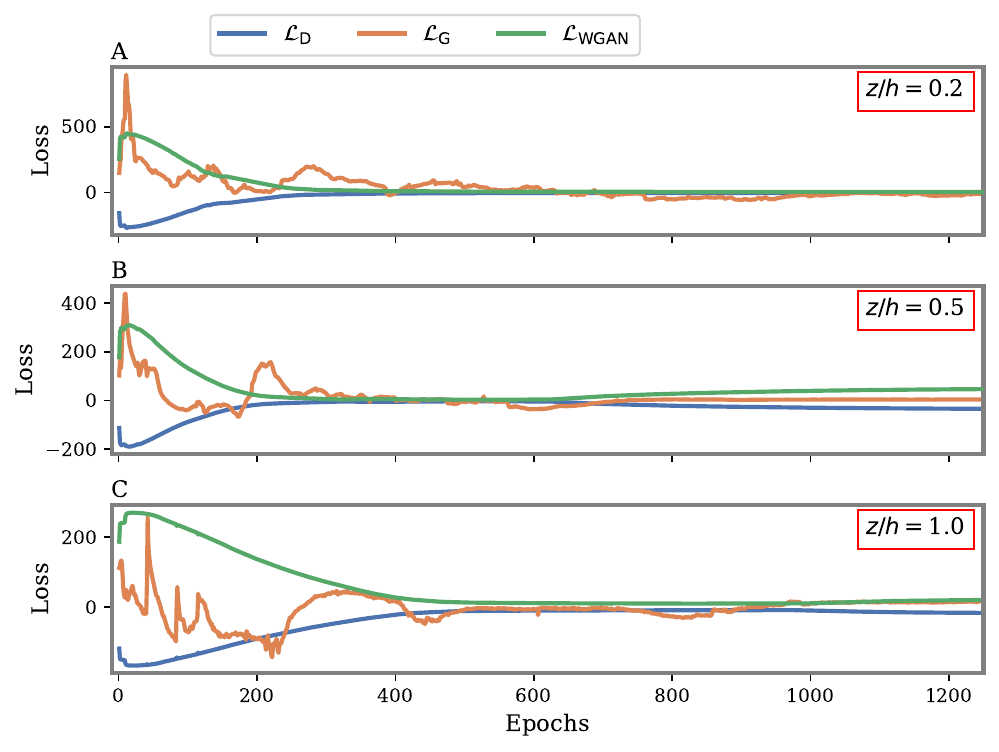}
\caption{Training loss of the three neural networks at height $z/h=0.2$ (A), $z/h=0.5$ (B) and $z/h=1.0$ (C). The discriminator (blue) and generator (orange) loss $\mathcal{L}_{\rm D}$ and $\mathcal{L}_{\rm G}$ converge towards individual values. The Wasserstein loss $\mathcal{L}_{\rm WGAN}$ (green) measures the quality of the GAN results. Hence the parameters of the GAN are saved at the minimum Wasserstein loss value.}
\label{fig:Appendix_Fig4}
\end{figure}

The GANs were trained for $2000$ epochs using the Adaptive moments (Adam) optimizer~\cite{Kingma2017}. An early stopping routine ceased training after $300$ epochs of no improvement of the Wasserstein metric $\mathcal{L}_{\rm WGAN}$. Each run was performed on two \textit{NVIDIA A100 Tensor-Core GPUs} and took up to 24 h to finish. The choice of GAN hyperparameters for the three planes is listed in Tab. \ref{tab:network_training}. Figure \ref{fig:Appendix_Fig4} shows additionally the losses for both, the generator and the discriminator, together with the Wasserstein loss for all three GANs.
%-------------------------------------------
\begin{table}[!htbp]
\centering
%\begin{tabular*}{\hsize}{@{\extracolsep{\fill}}ccccc}
\renewcommand{\arraystretch}{1.5}
\begin{tabular}{ccccc}
\hline
         plane $\tilde{z}$& learning rate & batch size & $N_{\rm D}/N_{\rm G}$ & $\lambda$\\
\hline
         $0.2$& $2\cdot 10^{-5}$ & $64$ & $12$ & $11$\\
         $0.5$& $2\cdot 10^{-5}$ & $64$ & $11$ & $10$\\
         $1.0$& $5\cdot 10^{-5}$ & $64$ & $90$ & $10$\\
\hline
\end{tabular}
    \caption{Neural network hyperparameters. The learning rate regulates the speed of the gradient descent update step. The batch size refers to the number of training examples processed together in a single forward and backward pass. The parameter $N_{\rm D}/N_{\rm G}$ signifies the degree to which the critic network undergoes additional training compared to the generator network. The gradient penalty parameter $\lambda$ controls the strength of the 1-Lipschitz condition in \eqref{eq:ML1}.}
    \label{tab:network_training}
\end{table}
%-------------------------------------------
%-------------------------------------------

\section{Additional results}
\begin{figure}[!htbp]
\includegraphics[width=\textwidth]{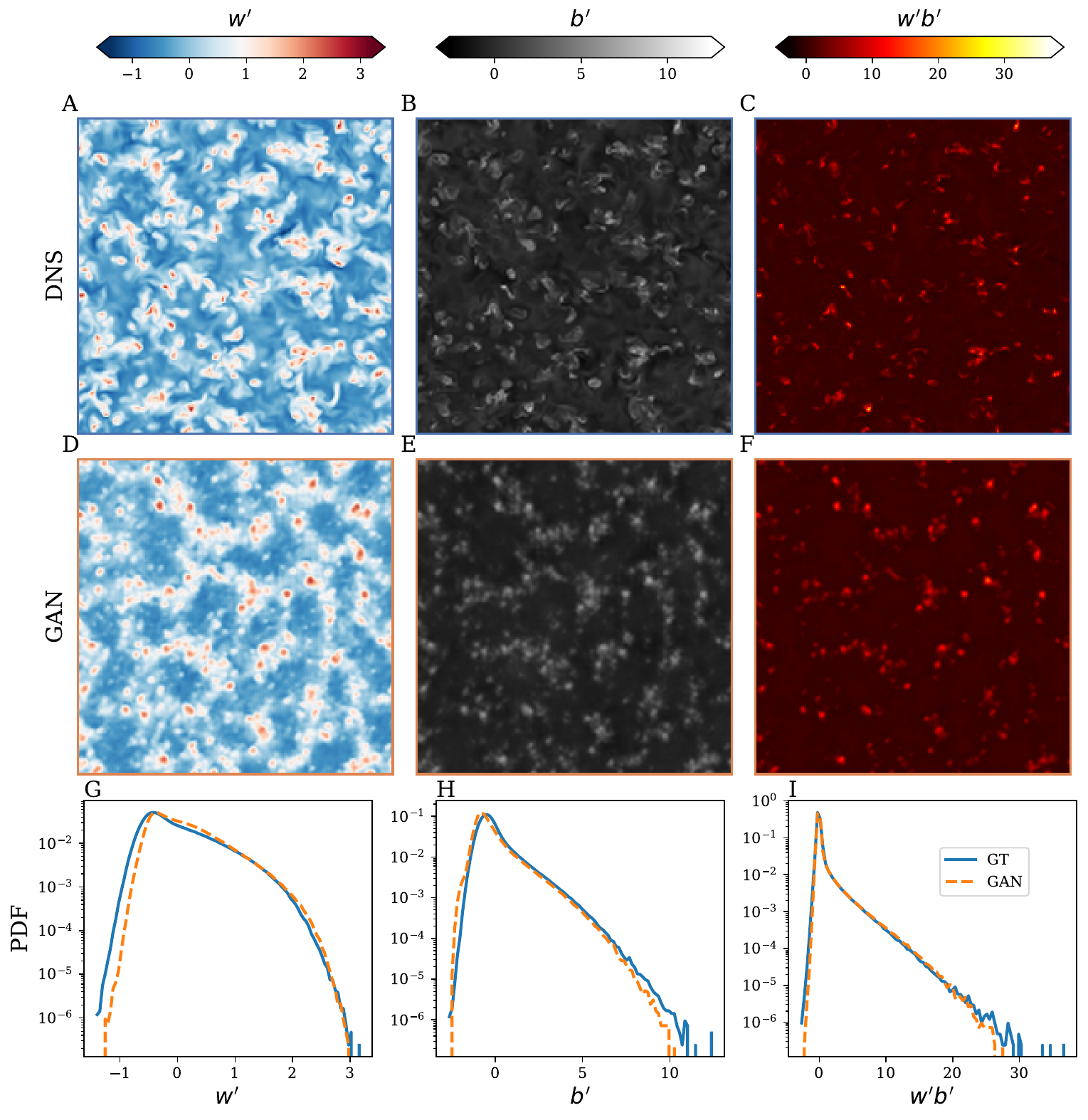}
\caption{Horizontal slices at $h/L_0=34.41$ in plane $z=0.5h = 0.43$. Compare to Fig. 4 in the main text.}
\label{fig:Appendix_Fig5}
\end{figure}

\begin{figure}[!htbp]
\includegraphics[width=\textwidth]{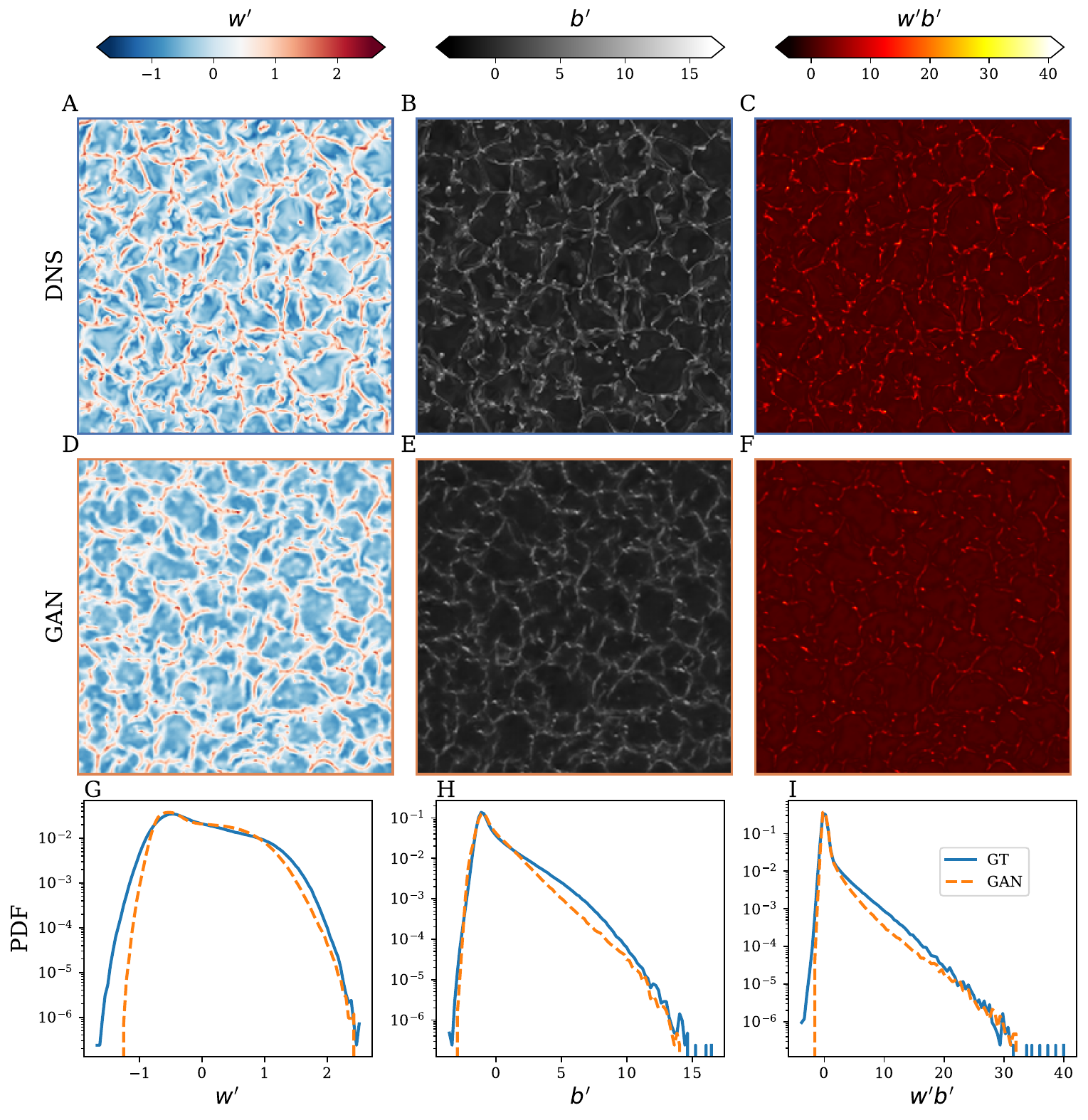}
\caption{Horizontal slices at $h/L_0=48.41$ in plane $z=0.2h = 0.233$. Compare to Fig. 4 in the main text.}
\label{fig:Appendix_Fig6}
\end{figure}

\begin{figure}[!htbp]
\includegraphics[width=\textwidth]{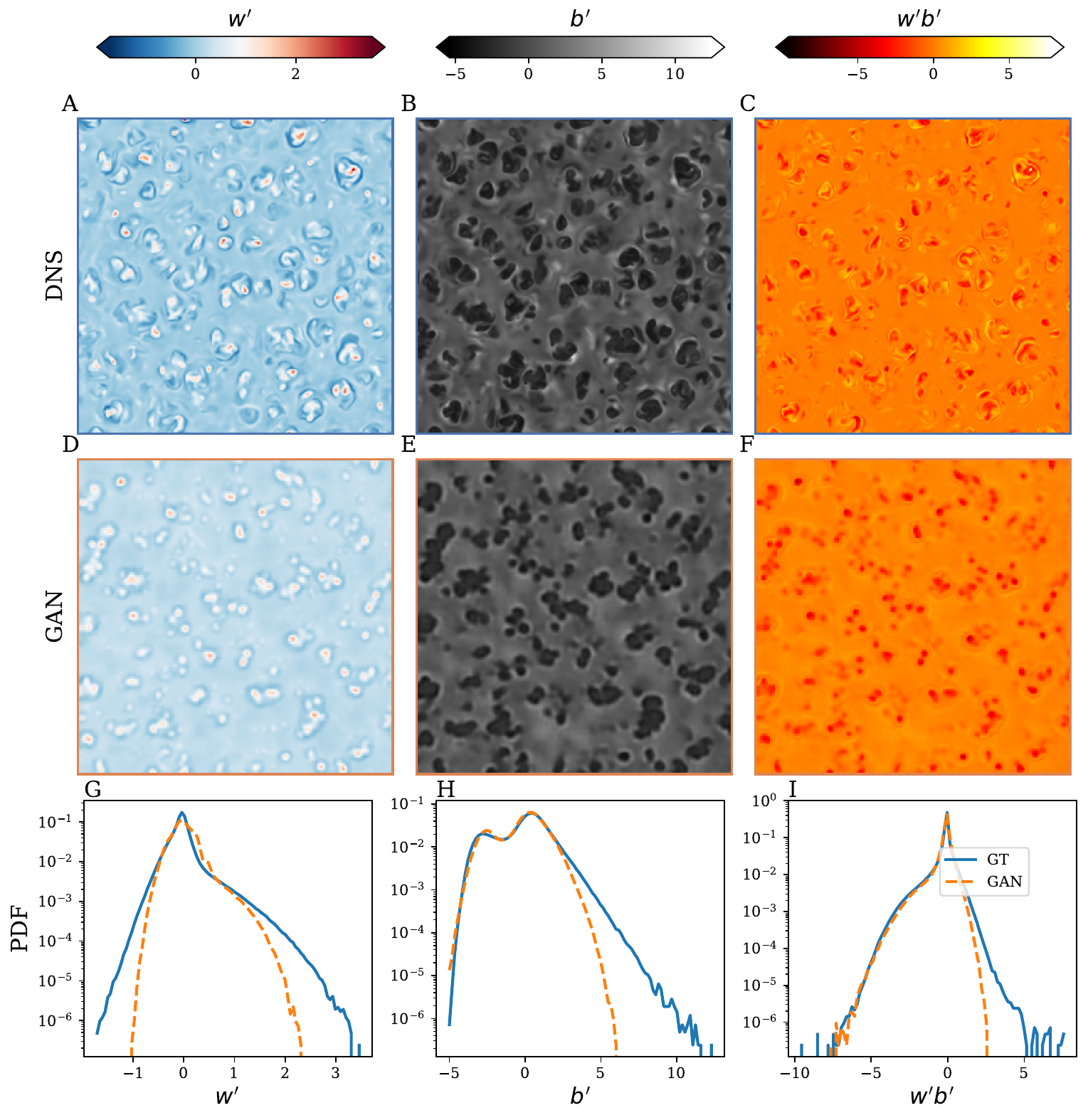}
\caption{Horizontal slices at $h/L_0=48.41$ in plane $z=1.0h = 1.167$. Compare to Fig. 4 in the main text.}
\label{fig:Appendix_Fig7}
\end{figure}

In this section, we supply further results that supplement the discussion of the main text. Supporting Figs. \ref{fig:Appendix_Fig5}, \ref{fig:Appendix_Fig6}, and \ref{fig:Appendix_Fig7} compare output of the GAN with the ground truth from the direct numerical simulations at different vertical positions $z$.

%----------------
% Bibliography
%----------------

\end{document}